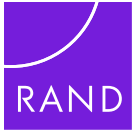

GOPAL P. SARMA, SUNNY D. BHATT, MICHAEL JACOB, RACHEL STERATORE

# Artificial General Intelligence Forecasting and Scenario Analysis

State of the Field, Methodological Gaps, and Strategic Implications

For more information on this publication, visit **www.rand.org/t/RRA4692-1**.

**About RAND**

RAND is a research organization that develops solutions to public policy challenges to help make communities throughout the world safer and more secure, healthier and more prosperous. RAND is nonprofit, nonpartisan, and committed to the public interest. To learn more about RAND, visit www.rand.org.

**Research Integrity**

Our mission to help improve policy and decisionmaking through research and analysis is enabled through our core values of quality and objectivity and our unwavering commitment to the highest level of integrity and ethical behavior. To help ensure our research and analysis are rigorous, objective, and nonpartisan, we subject our research publications to a robust and exacting quality-assurance process; avoid both the appearance and reality of financial and other conflicts of interest through staff training, project screening, and a policy of mandatory disclosure; and pursue transparency in our research engagements through our commitment to the open publication of our research findings and recommendations, disclosure of the source of funding of published research, and policies to ensure intellectual independence. For more information, visit www.rand.org/about/research-integrity.

RAND's publications do not necessarily reflect the opinions of its research clients and sponsors.

# About This Report

In this report, we review the current state of methodologies to forecast the arrival of artificial general intelligence, assess their reliability, and analyze the implications for strategy and policy. We synthesize diverse forecasting approaches, document significant limitations in existing methods, and propose a research agenda for developing more-robust forecasting infrastructure. The report does not endorse a specific forecast or scenario but rather provides a framework for interpreting forecasts under conditions of deep uncertainty.

We experimented with an iterative approach to human and artificial intelligence collaboration for this report. The primary drafting of the text was performed by large language models (GPT 5.1, Gemini 3 Pro, and Claude 4.5 Opus), with human researchers providing direction, peer review, fact-checking, and revision.

## Center on AI, Security, and Technology

RAND Global and Emerging Risks is a division of RAND that delivers rigorous and objective public policy research on the most consequential challenges to civilization and global security. This work was undertaken by the division's Center on AI, Security, and Technology, which aims to examine the opportunities and risks of rapid technological change, focusing on artificial intelligence, security, and biotechnology. For more information, contact cast@rand.org.

## Funding

This research was independently initiated and conducted within the Center on AI, Security, and Technology using income from operations and gifts and grants from philanthropic supporters. A complete list of donors and funders is available at www.rand.org/CAST. RAND clients, donors, and grantors have no influence over research findings or recommendations.

## Acknowledgments

This report went through numerous iterations, and many people contributed substantially to improving it. We thank several people whose valuable feedback underpins this effort: Zara Abdurahaman, Ben Boudreaux, Matan Chorev, Beba Cibralic, Ajeya Cotra, Forrest Crawford, Richard Danzig, Quentin Hodgson, Will Hurd, Brian Jackson, Kyle Kilian, Daniel Kokotajlo, Greg McKelvey, Nelson Lim, Jim Mitre, Arvind Narayanan, Joel Predd, William Regli, Stefanie Tompkins, and Bill Zito. We also thank our communications analyst Barbara Bicksler, the research editing and publications team (including Sean Mintus, Allison Kerns, and Brian Dau), and the Global and Emerging Risks quality assurance team (including Gary Cecchine, Jason Etchegaray, and Paige Smith) for helping us produce and deliver this timely report.

# Summary

Over the past five years, expert forecasts for achieving artificial general intelligence (AGI)—commonly referred to as *timelines* in artificial intelligence (AI) discourse—have shifted substantially from mid-century toward the near term. Prediction markets and compute-centric models (which project using trends in processing power, hardware investment, and the computational capacity used to train AI systems) now place central estimates in the 2030s, while the most recent (as of this writing in early 2026) large-scale expert survey puts the median at 2047 for high-level machine intelligence (HLMI)—a 13-year shift from a survey conducted just one year earlier. The same survey's "full automation of labor" estimate, which requires not just capability but also economic deployment, shifted from 2164 to 2116 (Grace et al., 2025).

We define AGI primarily as systems capable of performing most economically valuable work at or above human level across a wide range of domains—similar to the framing in OpenAI's founding charter and related (though not identical) to the HLMI construct used in expert surveys (defined as machines outperforming humans at every task). This definition of AGI emphasizes measurable task performance rather than autonomous goal-pursuit or economic transformation, although we discuss in Chapter 2 how alternative framings lead to different forecasts and strategic implications. When specific forecasts use alternative definitions, we highlight the distinctions.

In this report, we synthesize diverse AGI forecasting methodologies—including expert surveys, prediction markets, compute-centric models, and scenario analysis—to assess their reliability, identify the sources of expert disagreement, and develop decision frameworks for decisionmakers navigating uncertainty about both the timing and nature of advanced AI capabilities.

## Key Findings

**AGI timeline estimates have shifted earlier across methods.** Multiple independent forecasting approaches—expert surveys, prediction markets, and compute-centric models—show directional movement toward earlier AGI arrival dates. Although individual forecasters occasionally revise estimates into the future, the consistency of this shift across independent methods strengthens the signal; however, all methods share significant limitations.[1]

**Forecasting infrastructure is immature.** The field lacks resolved forecasts for calibration; benchmarks resistant to saturation and gaming; continuous, real-time insight into model capabilities; and independent validation of influential models. Decisionmakers are making decisions based on methodologies that are in nascent stages of development.

**Definitional ambiguity drives some, but not all, disagreement.** Much apparent disagreement reflects different definitions of AGI and different targets (technical capability vs. operational deployment vs. societal transformation). However, substantial disagreement remains even when definitions and information are held constant: People with similar training, working in the same organizations and looking at the same data, often reach very different conclusions about timelines and risk.

---

[1]  Timeline compression is also directionally consistent across different AGI definitions (e.g., HLMI, full automation of labor [FAOL], transformative AI), though the magnitude of compression varies substantially. In the Grace et al. (2025) survey, median estimates for HLMI compressed by 13 years relative to 2022, while FAOL compressed by 48 years. Notably, absolute timeline estimates differ enormously by definition—the 2023 median for FAOL (2116) is 69 years later than the median for HLMI (2047)—and framing effects within the same definition can shift estimates by ten to 15 years. Timeline estimates based on capability thresholds (e.g., in Metaculus resolution criteria for "weak" and "strong" AGI) have also shifted from 50 years to five years from 2025, even for resolutions that include robotics thresholds. The directional signal is robust, but precise timeline estimates remain highly sensitive to definitional and methodological choices.

AGI forecasts provide two kinds of value. First, some forecasting inputs carry genuine predictive signal that can inform near-term action: Compute[2] trends follow from hardware economics, capital investment reflects informed bets by actors with skin in the game, and scaling laws have shown empirical regularity. Second, the process of forecasting synthesizes disparate information in ways that clarify options for longer-term preparation. In this sense, forecasts function like scenarios in traditional national security analysis. Defense planners may debate the probability of specific contingencies, but they do not wait for consensus on those probabilities before developing response options; they identify scenarios that are plausible, consequential, and demanding of preparation and plan accordingly.

This framing suggests that the policy question is not "when will AGI arrive?" but "how should we prepare for a range of possible AI futures?" Effective strategy under such uncertainty requires three qualities: *flexibility* to pursue different objectives as circumstances evolve, *adaptiveness* to respond to unanticipated developments, and *robustness* to shocks. Given compressed and uncertain timelines, flexibility and adaptiveness are particularly critical. Static investments, however sound, will not be sufficient if the pace of development accelerates beyond current expectations.

Moreover, forecasts are one input into strategic decisions, not the sole determinant. Decisionmakers also bring values and objectives (what outcomes matter most), a space of possible options (what actions are feasible), constraints (fiscal, political, institutional), and risk tolerances (how to weigh uncertain harms against uncertain benefits). Two decisionmakers who accept identical forecasts may reach different conclusions: A government official and a venture capitalist may weigh security risks against innovation benefits differently; a frontier laboratory leader and a research funder may face different constraints on how quickly they can act. This report aims to improve the quality of forecasting as an input—clarifying what existing methods can and cannot tell us, identifying key uncertainties, and proposing investments that would strengthen the evidentiary base. It does not prescribe which values to prioritize or which options to pursue but instead aims to catalyze more-informed deliberation.

It is important to note that some credible forecasters—including leaders at frontier AI laboratories—place AGI arrival within the next one to four years. If they are correct, much of the institutional preparation this report describes would need to be radically compressed or may be overtaken by events entirely. This possibility is not an argument against preparation; it is an argument for urgency.

## Recommendations

The key findings above point to genuine uncertainty that may not resolve before consequential decisions must be made. The recommendations that follow span multiple actors. The first three address how decisionmakers should use forecasts and are directed primarily, but not exclusively, to U.S. government policymakers: executive branch agencies, Congress, and the national security community. The remaining five focus on strengthening the forecasting infrastructure that informs those decisions, addressing a broader set of actors: Research funders (National Science Foundation, Defense Advanced Research Projects Agency [DARPA], Intelligence Advanced Research Projects Activity [IARPA], private philanthropy) can support methodological diversity and validation infrastructure; academic researchers can provide independent stress-testing and develop novel approaches; frontier AI laboratories can contribute to capability evaluation and internal monitoring of AI-assisted research and development; and forecasting platforms and research organizations (Epoch AI, Model Evaluation and Threat Research [METR], Metaculus) can expand tracking infrastructure. Some investments would build on existing platforms (e.g., expanding Epoch AI's empirical tracking, supporting forecasting tournaments on intermediate milestones); others involve building government technical expertise to assess developments and partner effectively with industry.

---

[2]   Throughout this report, *compute* (informal usage) refers to computing resources: the processing power, hardware infrastructure, and computational capacity required to develop and run AI systems. The term emerged in the era of cloud computing as shorthand for what would more formally be called *computational resources*.

**Treat forecasts as scenario-structuring tools, not point estimates to optimize around.** Rather than debate precise probabilities, planners should focus on scenarios that are plausible, consequential, and challenging—particularly those that lack adequate preparation. Given the potential consequences, a Taiwan contingency or 9/11-style attack require comprehensive planning, even if the likelihood of such an event is seen as low. AGI forecasting should function similarly. Framing the issue as "how should we prepare for a range of possible AI futures?" shifts attention from prediction accuracy to decision-relevant preparation. Concretely, strategic and contingency planning processes should incorporate AI-related scenarios, including short-timeline trajectories, among the contingencies for which agencies develop response options.

**Build adaptive capacity with clear reassessment triggers.** Decisionmakers should develop contingency plans for near-term AGI scenarios, establish explicit triggers for reassessment, and match action timing to the type of evidence most relevant to each domain. For example, safety and security measures should be informed by capability developments and assessed in the context of whether those capabilities meaningfully alter existing risk landscapes, such as the attacker-defender balance in cybersecurity or the effectiveness of existing safeguards in biosecurity. Innovation investments should proceed continuously regardless of timeline beliefs. Workforce adjustments can respond to observed market effects, such as labor displacement and productivity shifts. Triggers should be concrete and tied to observable indicators: For example, when AI systems demonstrate the capacity to autonomously complete multiweek software engineering projects, or if the percentage of AI research substantially automated by AI systems exceeds 50 percent, assumptions underlying longer-timeline scenarios would require revision. Success means being positioned to respond effectively across a range of futures, and experience suggests that planning for more challenging or inconvenient scenarios often yields tools and responses that serve well in less demanding ones. This approach argues for guarding against the tendency to anchor planning on comfortable, seemingly probable futures.

**Make forecasts decision-relevant by linking them to strategic choices.** Decisionmakers have historically underused forecasts, in part because they do not illuminate how decisions affect outcomes. A forecast that there is a 20 percent chance of a given event tells a decisionmaker little about whether a particular course of action would shift that probability. To increase uptake, forecasts should be structured around conditional questions: How do projected timelines or capability trajectories change under different investment, strategic, or diplomatic scenarios? Integrating forecasting with scenario planning methods can help surface these contingencies. Institutionalizing this integration will likely require directed coordination at senior levels; experience suggests that forecast-informed planning rarely emerges through ad hoc processes alone.

The remaining recommendations focus on strengthening the forecasting infrastructure that informs these decisions.

**Invest in methodological diversity and novel approaches.** Current forecasting draws heavily on a narrow set of methods and disciplinary perspectives. Bringing in econometricians, cognitive scientists, historians of technology, and complex systems researchers could surface blind spots and challenge shared assumptions. Structured disagreement matters more than consensus. What decisionmakers need is clarity on whether experts diverge about empirical facts, theoretical models, or normative values. Agencies such as the National Science Foundation, DARPA, and IARPA are well-positioned to support this work.

**Institutionalize independent validation and stress-testing of influential models.** The compute-centric models and takeoff analyses that increasingly shape expectations and investment in AI deserve adversar-ial scrutiny comparable to that applied in climate modeling, intelligence forecasting, or macroeconomic projection—domains in which the stakes similarly justify systematic validation. This means red-teaming core assumptions, conducting transparent sensitivity analyses, and systematically comparing model predictions against outcomes as evidence accumulates. Forecasting institutions should be independent of the organizations whose futures they are predicting. Durable validation capacity should be built across multiple institutional homes—academia, civil society, government, and international partnerships—to ensure continuity and independence over time.

**Invest in independent, ongoing capability evaluation.** Current benchmarks degrade quickly—through contamination, saturation, and optimization pressure. These are not hypothetical concerns: Leading benchmarks often saturate within two years of release, and evidence of training data contamination has been documented across widely used evaluation suites. Rather than seeking a definitive benchmark, decisionmakers should support evaluation infrastructure that continuously develops new assessments, maintains held-out test sets, and tracks real-world task performance over time. Promising approaches include benchmarks that refresh monthly with novel questions, dynamic test generation that creates evaluation instances at runtime, and held-out evaluation sets maintained under strict access controls. Sustained investment in this function should be housed outside frontier laboratories. Organizations such as METR, Epoch AI, Apollo Research, and RAND perform aspects of this work, as do government bodies, including the United Kingdom's AI Security Institute and the U.S. Center for AI Standards and Innovation; scaling this capacity should be a priority for government funders (including the National Science Foundation and National Institute of Standards and Technology) and philanthropic funders alike.

**Develop monitoring infrastructure suited to compressed timelines.** If transformative capabilities emerge in the late 2020s, annual assessments will be too slow to inform decisions. Short-timeline scenarios require higher-frequency, more heterogeneous approaches—closer to quantitative finance than traditional technology forecasting. This means tracking multiple leading indicators in real time, rapidly updating as evidence arrives, and developing monitoring systems that can detect discontinuities rather than just extrapolate trends. Relevant indicators might include time-horizon metrics tracking the duration of tasks AI agents can complete autonomously, performance trajectories on privately held benchmarks, and qualitative assessments from AI researchers on the degree of AI assistance in their work. Triangulation across multiple imperfect signals matters more than any single, definitive metric.

**Strengthen internal monitoring of AI automation of AI research and development.** The degree to which AI systems are accelerating AI research is the leading indicator most relevant to rapid capability gains and potential discontinuities. Frontier laboratories are best positioned to track this internally; they should develop and improve standardized monitoring systems now using formats that could support broader information-sharing if coordination becomes necessary. This monitoring presents a collective action problem: Individual labs may be reluctant to develop or share such metrics unilaterally, given competitive pressures and concerns about regulatory or reputational consequences. Industry coordination, potentially formalized through voluntary commitments analogous to responsible scaling policies or through government-facilitated information-sharing agreements, may be necessary to establish baseline measurement practices. The alternative—external observers attempting to infer automation levels from indirect signals—would be far less reliable.

## Summary Reference

# Contents

# Figure and Tables

## Figure

## Tables

# Introduction

Artificial intelligence (AI) systems are increasingly embedded in critical infrastructure, from search and productivity tools to logistics and national security applications. From government officials setting national policy, to investors allocating capital, to laboratory leaders planning research agendas, the strategic challenge extends beyond managing existing systems to anticipating the trajectory ahead. A central question has emerged: When might AI systems achieve capabilities comparable to human-level general intelligence? Relatedly, what would such a development imply for decisionmakers?

This question (the timing of *artificial general intelligence*, or AGI) has moved from speculative discussion to active strategic and policy concern. Multiple signals suggest accelerating progress: empirical regularities in how model capabilities scale with computational resources (Kaplan et al., 2020; Hoffmann et al., 2022), unprecedented capital investment in AI infrastructure (Williams, 2025; Levy, 2025; Sevilla and Roldán, 2024), and a marked shift in expert expectations toward earlier arrival dates (Grace et al., 2025; Todd, 2025).

However, decisionmakers interpreting the available evidence face a difficult situation. Forecasting methods are nascent, with limited track records and significant methodological debates (Armstrong, Sotala, and Ó hÉigeartaigh, 2014; Steinhardt, 2023). Expert opinion, although shifting toward earlier AGI timelines, remains divided (Müller and Bostrom, 2016; Grace et al., 2025). The phenomenon being predicted (AGI) is unprecedented, making historical analogies of limited value. Yet as Danzig (2011) and Chorev and Predd (2025) argue, national security decisions routinely require action under conditions of fundamental uncertainty about technological trajectories; the challenge is not to eliminate uncertainty but to develop strategies that are robust to it.

## Report Objectives

In this report, we do not attempt to provide a definitive timeline to AGI. Given current uncertainties, such precision would be misleading. Instead, we aim to accomplish the following:

- **Define the target.** Clarify what AGI means across different definitions, why definitional choices matter for strategy and policy, and how capability demonstration, deployment, and societal transformation may occur on different timelines (Chapter 2).
- **Synthesize current forecasts.** Present the range of predictions from major forecasting approaches, including their methodological foundations and key assumptions (Chapter 3).
- **Explain disagreement.** Identify the specific empirical and conceptual questions that drive divergent predictions (the "cruxes"), distinguishing substantive disputes from definitional confusion (Chapter 4).
- **Assess reliability.** Evaluate the methodological foundations and limitations of current forecasting approaches, documenting where confidence is warranted and where it is not (Chapter 5).
- **Develop strategic and policy implications.** Provide a framework for decisionmaking under uncertainty, distinguishing appropriate decisionmaking triggers and identifying indicators for ongoing assessment (Chapter 6).

- **Propose research priorities.** Outline specific investments that would improve forecasting infrastructure and provide decisionmakers with better decision-relevant information (Chapter 7).

This analysis largely assumes that the current trajectory of investment, scaling, and improvement will persist, even if at varying rates. This assumption is not certain. AI has experienced *winters* before, a time when research advancements decrease during a dormant period, and the conditions for another winter remain plausible (technical plateau, compute bottlenecks, credit tightening, public backlash, or regulatory intervention). A sustained stagnation scenario would substantially alter the strategic landscape: Questions about AGI timelines would recede, and analysis would shift toward the dynamics of the broader U.S. research and development (R&D) ecosystem, innovation policy, and technological competitiveness. The economics literature on technological change—including work on technological costs and forecasting (Farmer and Lafond, 2016), cycles of innovation and deployment (Perez, 2002), and hype-disillusionment dynamics (Fenn and Raskino, 2008)—would become more directly relevant than the AGI-specific forecasting methods reviewed here. We do not develop that alternative scenario in detail.

Although we draw on global forecasting research and consider AI development as an international phenomenon, our recommendations focus primarily on U.S. decisionmakers and institutions. Forecasting relative international AI capabilities and comparative analysis of national strategies are beyond the scope of this report. The White House AI Action Plan released in July 2025—which establishes federal policy across the three pillars of innovation, infrastructure, and international diplomacy—provides the immediate institutional context for several of our recommendations, particularly those addressing evaluation infrastructure, workforce impacts, and national security risk assessment (White House, 2025).

## Methodology

This report was developed through iterative human-AI collaboration. Large language models (LLMs; initially GPT-5.1 and Gemini 3 Pro and later Claude Opus 4.5) generated draft syntheses based on structured outlines and conceptual direction provided by the lead researcher. We reviewed these drafts for accuracy, verified citations against original sources, supplemented the analysis with recent (late 2025 and early 2026) publications, and iteratively refined the content through multiple revision cycles. The approach is described in detail below.

This report is not a systematic review. The forecasting literature it synthesizes exists predominantly in nontraditional venues—research organization blogs, online discussion forums, prediction market platforms, and independently published analyses—rather than in indexed academic journals. These are nonetheless highly influential sources: Models such as Biological Anchors (Cotra, 2020; Cotra, 2022) and Davidson's takeoff framework (Davidson, 2023) are shaping expectations about AI timelines across research, industry, and government, and the researchers producing them represent some of the deepest thinkers on these questions. Our approach was to leverage domain expertise to ensure thorough coverage of the primary frameworks and themes driving discussion in the field and then to broaden from there into adjacent debates and emerging evidence. The goal was a broad and accessible synthesis that would allow researchers, analysts, and decisionmakers to orient to the landscape of AGI forecasting and the key sources of disagreement within it. We hope this work serves as a foundation; a more systematic treatment of this literature would be a valuable contribution.

## Research Questions

AI forecasting has become simultaneously more urgent and more contested, with decisionmakers requiring synthesis across diverse and often contradictory prediction methods. Our analysis was structured around four core research questions that connect methodological assessment to practical decisionmaking needs:

- **Synthesis across methodological diversity.** How can diverse approaches to AGI timeline estimation—expert surveys, crowd platforms, compute models, and scenario analysis—be meaningfully integrated despite their different definitions, assumptions, and outputs?
- **Reliability assessment.** What are the empirical foundations, track records, and systematic biases of different forecasting methods, and how should these characteristics inform the relative weighting of evidence?
- **Decisionmaking under disagreement.** Given fundamental disputes about whether current AI approaches can achieve AGI, what robust frameworks can guide action when experts disagree not just on timing but also on feasibility?
- **Methodological gaps.** What critical deficiencies in current forecasting methods limit their utility for strategic planning, and what specific research investments could improve the quality of decision-relevant information?

These questions reflect the report's dual purpose: providing immediate guidance for decisionmakers who must act despite uncertainty while identifying longer-term improvements to forecasting infrastructure that could reduce that uncertainty over time.

## Human-AI Collaborative Approach

This report was developed through iterative human-AI collaboration between October and December 2025. The methodology, depicted in Figure 1.1, reflects an emerging approach to policy research, one that leverages the synthesis capabilities of LLMs while maintaining human direction and quality control. As AI capabilities evolve rapidly, the methods for productively incorporating them into research workflows are themselves evolving; this report represents one approach, not a settled template.

The process worked as follows. The lead researcher provided structured outlines and conceptual direction based on domain expertise in AI governance and forecasting methodologies. LLMs from several frontier AI vendors generated draft syntheses, drawing on knowledge from their training sets—initially GPT-5.1 and Gemini 3 Pro and eventually including Claude Opus 4.5. The research team reviewed these drafts for accuracy, verified citations against original sources, supplemented the analysis with recent publications, and iteratively refined the content through multiple revision cycles, with the aid of the LLMs, as Figure 1.1 depicts in the second stage. Revised drafts were further reviewed and refined, with emphasis at this stage on human review, leading to the final research report. Although human judgment guided all substantive decisions, the prose was primarily machine-generated.

Figure 1.1 presents a simplified representation of this workflow. The actual process was substantially more complex, involving feedback loops within feedback loops at every stage. Initial generation involved multiple rounds of prompting, refinement, and comparison across models. The review and revision stage shown as a single cycle in the figure actually comprised dozens of iterative exchanges: Human reviewers identified specific issues (citation errors, unclear arguments, missing context), AI systems proposed revisions, humans assessed those revisions and often requested further refinement, and this process repeated until quality standards were met. No systematic pattern suggested that AI-generated errors biased the analysis in a particular direction, though we remained alert to the possibility of shared blind spots across models. Even "final" stages involved returning to AI systems for localized improvements, fact-checking

**FIGURE 1.1**

**Illustrative AI-Driven Research Report Process: Collaborative Human-AI Methodology**

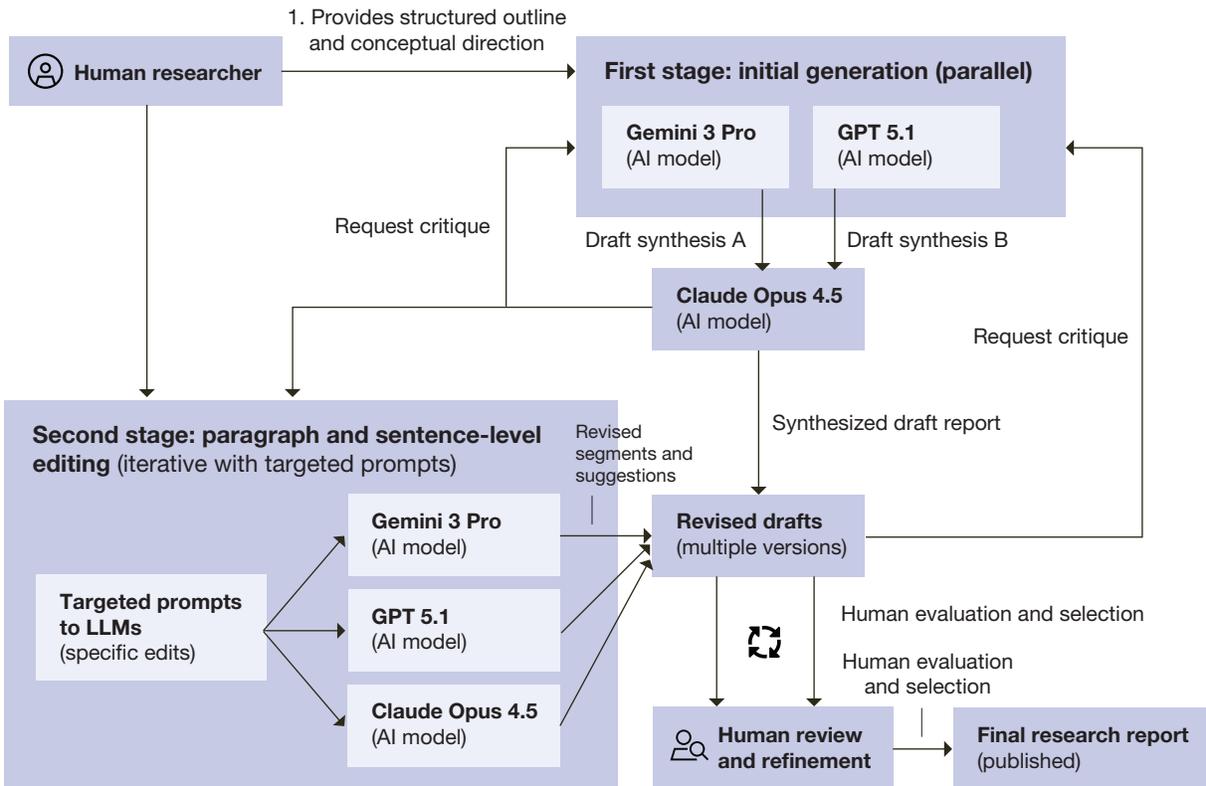

assistance, and incorporation of late-breaking developments. The figure's linear progression obscures the reality that human judgment and machine assistance were deeply interwoven throughout, with neither operating in isolation for any extended period.

In addition to the ongoing validation of references throughout the drafting and revision process and our ordinary institutional quality assurance review, a final dedicated citation verification pass was conducted across the full manuscript, with each coauthor independently reviewing every reference using a combination of LLM-assisted and human-only methods. We take responsibility for any errors that remain but are confident that they do not materially affect the report's conclusions.

This division of labor played to the respective strengths of human researchers and AI systems. The AI systems provided broad coverage of a large literature and rapid generation of structured prose. Human researchers provided strategic direction, domain judgment, fact-checking, and quality control—functions that remain essential, given current AI limitations related to factual accuracy and source verification. We validated factual claims and citation accuracy, made judgments on topical relevance and recommendation priorities, and ensured logical coherence throughout. Human review identified the following recurring issues in AI-generated drafts that required correction:

- Citation errors. AI systems occasionally generated plausible-sounding but incorrect citations, including wrong publication years, misattributed quotes, and references to the wrong version of papers or papers that did not exist. All citations were verified against original sources.
- Factual inaccuracies. Draft content sometimes included outdated information, conflated distinct concepts, or overstated the confidence of source claims.

- Structural issues. AI-generated drafts occasionally repeated points across sections, failed to maintain consistent terminology, or introduced claims without adequate support. These issues required reorganization and consolidation during revision.

The resulting analysis should be understood as a structured synthesis of publicly available AI forecasting literature rather than original empirical research. It provides a framework for understanding disagreements and uncertainties rather than resolving them. With this scope, we draw primarily from publicly available forecasting research, expert surveys, and prediction market data; we do not draw from classified assessments, proprietary information from AI laboratories, or primary data collection.

Use of this human-AI collaboration enabled a small team to produce more comprehensive and more rigorous synthesis more quickly, which is critical to keeping pace with a rapidly evolving and highly contested policy and technological field. Whether this represents a model for future work will depend on continued improvement in AI capabilities and the development of best practices that are, at present, being discovered through experimentation.

Much of the iterative workflow described above (the cycling between generation, review, revision, and fact-checking across multiple models) required sustained human orchestration. Multi-agent systems capable of managing these feedback loops autonomously are developing rapidly, and it is likely that future iterations of this methodology will automate significant portions of the process. This also means that exact replication of the workflow described here may not be feasible or desirable: The models used have already been updated or superseded. Prompting strategies that were effective in late 2025 may not produce comparable results on newer systems, and the manual coordination that characterized this effort is already giving way to more-automated approaches. We document the process as practiced, recognizing that the tools and methods are evolving faster than any fixed protocol can capture.

The pace of development during the writing of this report illustrates the challenge of analyzing a field that moves faster than the publication cycle. Between November 17 and December 11, 2025, four frontier laboratories released major new models within a 25-day span, each briefly claiming top positions on capability benchmarks before being overtaken by the next.[1] The International AI Safety Report, published in January 2025, required two key updates (in October and November 2025) because capabilities advanced faster than its annual reporting cycle could accommodate; a full second edition followed in February 2026 (Bengio, Clare, et al., 2025; Bengio, Mindermann, et al., 2025; Bengio et al., 2026). The Forecasting Research Institute published the first results from its Longitudinal Expert AI Panel in November 2025 (Murphy et al., 2025).

The pace has not slowed during the peer-review process. On February 5, 2026, Anthropic and OpenAI released new frontier models within minutes of each other (Claude Opus 4.6 and GPT-5.3-Codex); OpenAI described its model as "our first model that was instrumental in creating itself" (Anthropic, 2026; OpenAI, 2026). METR's December 2025 evaluation of Claude Opus 4.5 estimated a 50 percent task-completion time horizon of nearly five hours, above the exponential trend (Kwa et al., 2025; Model Evaluation and Threat Research [METR], 2026). In late January, the Moltbook episode (involving a viral social network populated by more than 1 million AI agents) captured public attention before security researchers revealed exposed databases and largely human-directed activity (Dellinger, 2026; Nagli, 2026), illustrating both how rapidly agent capabilities are advancing and how easily the discourse around them can outrun the evidence. Each of these developments arrived while this report was in preparation or review, reinforcing the central finding that the window for building institutional capacity to respond is narrowing.

---

[1] The four models were xAI's Grok 4.1, Google's Gemini 3, Anthropic's Claude Opus 4.5, and OpenAI's GPT-5.2 (xAI, 2025; Pichai, Hassabis, and Kavukcuoglu, 2025; Anthropic, 2025; OpenAI, 2025; Arena, undated).

# Chapter 1 References

# Defining the Target

Before assessing forecasts, it is essential to clarify what is being predicted. Although there are substantive empirical disputes worth examining, much apparent disagreement about AGI timelines reflects definitional differences. Decisionmakers should understand that AGI is not a single, well-defined threshold but a family of related concepts with different strategic implications.

## Common Definitions

Several definitions of AGI appear in forecasting literature and industry discourse.

**Economic task completion** indicates systems that are capable of performing "most economically valuable work" at or above human level. This definition, used by such organizations as OpenAI in its founding charter, emphasizes economic impact and measurable task performance but leaves ambiguity about which tasks qualify and what "most" means (OpenAI, undated).

**High-level machine intelligence (HLMI)** indicates systems that can "carry out most human professions at least as well as a typical human." This definition, common in expert surveys, is similar to economic task completion but framed around occupational categories rather than economic value (Müller and Bostrom, 2016).

**Top human-Expert Dominating AI (TED-AI)** indicates systems that match or exceed the best human professionals at virtually all cognitive tasks, including meta-tasks, such as flexibly adapting to new situations and learning new skills. This definition, introduced in forecasting models developed by Greenblatt (2024) and adopted by Kokotajlo et al. (2025), sets a higher bar than HLMI by comparing AI with domain experts rather than typical practitioners (i.e., the capacity to automate almost all remote cognitive work, given adequate resources).

**Cognitive capability equivalence** indicates systems that match the cognitive versatility and proficiency of a well-educated adult across core cognitive domains. Hendrycks et al. (2025) ground this approach in Cattell-Horn-Carroll theory, the most empirically validated model of human cognition, decomposing general intelligence into ten domains—including reasoning, memory, and perception—and adapting established psychometric batteries to evaluate AI. This framework yields quantifiable AGI scores (e.g., GPT-4 at 27 percent, GPT-5 at 57 percent), revealing that current systems have highly uneven cognitive profiles with critical deficits in foundational areas, such as long-term memory.

**Autonomous goal-pursuit** indicates systems that are capable of independently formulating and pursuing complex goals across diverse domains without human specification of methods. This definition emphasizes agency and generalization rather than task performance. Google DeepMind's "Levels of AGI" framework incorporates this dimension, describing the progression of AI capability along an autonomy axis (Morris et al., 2024).

**Recursive self-improvement** indicates systems that are capable of automating all or nearly all of the AI R&D process, thereby accelerating capability gains. This definition focuses on a specific capability with sig-nificant safety implications: The speedup factor could be dramatic if AI systems perform the bulk of AI R&D, creating a feedback loop that compresses subsequent development timelines. Existing AI systems already contribute to their own improvement in limited ways (e.g., assisting with code generation and experiment design or providing training data for future models), but the threshold of concern is comprehensive automation of the R&D pipeline. Anthropic's Responsible Scaling Policy identifies autonomous AI R&D as a key capability threshold warranting enhanced safety measures (Anthropic, 2023; Anthropic, 2025).

**Scientific research automation** indicates systems that are capable of conducting novel scientific research at or above human expert level, encompassing the full research pipeline from hypothesis generation through experimental design and execution. Karnofsky's concept of a "Process for Automating Scientific and Technological Advancement" was among the first to articulate this as a threshold for transformative AI (Karnofsky, 2021). The Nobel Turing Challenge proposed that AI capable of making discoveries worthy of a Nobel Prize would represent a key milestone (Kitano, 2016). More recently, FutureHouse, a nonprofit founded in 2023, explicitly adopted the concept of an AI Scientist as its development target: systems capable of autonomously generating hypotheses, synthesizing literature, and directing experiments (FutureHouse, undated; White and Rodriques, 2025).

**Transformative AI** indicates systems that cause very large economic or societal changes, originally defined by Karnofsky (2016) as AI that causes a change as large as or larger than the First Agricultural Revolution or Industrial Revolution and often operationalized as driving a tenfold acceleration in economic growth or automating a substantial fraction of cognitive labor. This definition, developed further in influential forecasting models (Cotra, 2020; Davidson, 2021), focuses on measurable macro-level impact rather than capability thresholds.

In this report, AGI refers primarily to human-level general capability (closer to economic task completion and HLMI) rather than the broader transformative framing. The following section introduces a framework for distinguishing these stages; where we discuss forecasting models that use transformative AI as their target, we note which stage they primarily address.

## The Three-Stage Framework: Capability, Deployment, Societal Transformation

As mentioned earlier, understanding AGI forecasts requires distinguishing what, precisely, is being predicted. Different forecasters target different phenomena. Some predict when systems will first demonstrate AGI-level performance, others when such systems will be widely deployed, and still others when societal transformation will become measurable. These distinctions matter because they imply different decisionmaking triggers and different lead times for response. The framework below makes these distinctions explicit.

A critical source of confusion in AGI forecasting is the conflation of technical capability with operational use and societal transformation. We distinguish among the following three stages:

- **Capability existence.** The first time an AI system with a given capability exists, likely within the company that developed it. Whether that capability can be definitively established through controlled evaluation is uncertain: Benchmarks can demonstrate narrow task performance, but the broad, real-world capabilities of greatest concern may be difficult to verify in laboratory settings.
- **Operational deployment.** The point at which those capabilities are integrated into widely used products or workflows but before broad societal effects materialize.

- **Societal transformation onset.** The point at which AI capabilities begin measurably affecting productivity, labor markets, national security, or geopolitical dynamics at a significant scale.[1]

Historical precedent from general-purpose technologies suggests that gaps between these stages can be substantial. Research on the electrification of U.S. manufacturing shows that sectorwide productivity gains lagged initial adoption by two to four decades (David, 1990; Fiszbein et al., 2020; Narayanan and Kapoor, 2025). However, the analogy has limits. Software-based technologies can distribute nearly instantaneously via cloud infrastructure, and AI adoption has been rapid: ChatGPT reached 100 million users within two months of launch (Hu, 2023; Berg and Ho, 2025). Moreover, sufficiently capable AI systems might compress these stages by strategically facilitating their own deployment. The three-stage framework remains useful for distinguishing what is being predicted, but decisionmakers should not assume that historical gaps between stages will hold for AI.

Computer scientists and AI researchers typically forecast capability existence. Economists typically forecast economic transformation. A researcher predicting "AGI by 2030" and an economist predicting "no major labor impact until 2045" may both be correct because they are predicting different phenomena.

For decisionmaking purposes, all three stages matter, but different strategic goals will emphasize different indicators depending on the domain. Safety and security measures should be informed by capability developments, as dangerous capabilities can be consequential before widespread deployment. Labor and education policies can track economic transformation, which provides more lead time for observation and adjustment. The appropriate trigger depends on the strategic objective and the costs of acting too early versus too late.

## Why Definitions Matter for Strategy and Policy

These definitions are not merely semantic. They imply different timelines, different decision-relevant triggers, and different risk profiles.

**Narrow economic capability may emerge before general autonomy.** A system that excels across many cognitive domains might qualify as AGI under economic definitions while lacking the autonomous goal-pursuit that raises safety concerns. Conversely, AI capabilities may remain jagged (exceptional in some domains while struggling in others) rather than smoothing into uniform general competence (Toner, 2025; Dell'Acqua et al., 2023; Danzig, 2025).

**Dangerous capabilities may precede AGI under any definition.** Systems that fall short of general intelligence might nonetheless pose significant risks in such domains as cybersecurity, biological research, and persuasion. This recognition underlies capability-based regulatory frameworks adopted by multiple frontier AI laboratories—including OpenAI's Preparedness Framework, Anthropic's AI Safety Levels, and Google DeepMind's Frontier Safety Framework—which trigger enhanced safety protocols based on specific dangerous capabilities rather than general intelligence thresholds (OpenAI, 2023; Anthropic, 2025; Dragan, King, and Dafoe, 2024).

---

[1] Some frameworks add a fourth, *technology-input* notion of AGI—e.g., defining thresholds in terms of training compute (floating-point operations per second [FLOPS] counts) or related hardware or energy inputs. This addition can be attractive for decisionmakers because inputs may be comparatively observable and administrable, but it is also crude: Input levels do not uniquely determine capability and can incentivize threshold-gaming, and key inputs are often proprietary or difficult to verify. For these reasons, we treat input-based thresholds as a supplementary policy heuristic rather than a core stage of the capability → deployment → transformation framework.

**Technical capability differs from widespread deployment.** A system that achieves AGI-level performance within the organization that developed it may not be widely available for some time because of legal, regulatory, or strategic constraints.

Wargaming exercises conducted in 2024 demonstrated this dynamic directly: Participants' policy recommendations varied dramatically depending on how they defined AGI and how much lead time they believed was available (Predd, Chessen, and Smith, 2025).

## Chapter 2 References

# Forecasting Methodologies

AGI forecasting employs several distinct methodological approaches. Most methods primarily forecast capability demonstration, with fewer directly targeting deployment or societal transformation; as discussed in the previous chapter, this mismatch underlies some apparent disagreements between technical and economic forecasts. This report focuses on quantitative and semi-quantitative forecasting methods: expert surveys, prediction markets, compute-centric models, and empirical trend extrapolation.

However, the broader ecosystem of future-oriented analysis extends further. Scenario-based planning, agent-based modeling of sociotechnical systems, and structured speculation—including science fiction—can complement these approaches. Science fiction in particular has demonstrated predictive value for discontinuous change: Concepts such as cyberspace, ubiquitous social media, and the attention economy were explored in fiction before becoming real-world manifestations and strategic concerns. These narrative and simulation-based methods may be especially useful for reasoning about qualitative breaks from trend rather than extrapolations and for helping decisionmakers develop intuition for possibilities that quantitative models struggle to capture.

## Expert Surveys and Elicitation

**Method.** This approach employs structured surveys asking AI researchers and related experts to estimate probabilities or dates for AGI-related milestones. Major surveys include Grace et al. (2018); Grace et al. (2025); Stein-Perlman, Weinstein-Raun, and Grace (2022); Müller and Bostrom (2016); and the Forecasting Research Institute's Longitudinal Expert AI Panel (Murphy et al., 2025).

**Key findings on AGI timelines.** A 2023 survey of AI researchers found a median estimate of a 50 percent probability of HLMI by 2047. This response demonstrated a 13-year shift from a similar survey in 2022, which had a median estimate of 2060 for HLMI (Stein-Perlman, Weinstein-Raun, and Grace, 2022; Grace et al., 2025). Experts surveyed in the Forecasting Research Institute's Longitudinal Expert AI Panel expected sizable societal effects of AI by 2040, with the median expert predicting that AI will be comparable to a "technology of the century" (Murphy et al., 2025).

**Strengths.** Surveys capture the beliefs of domain experts who have direct knowledge of technical progress and challenges. Asking similar questions at different points in time can capture expert updates based on significant real-world events (such as the release of GPT-3). Large sample sizes can reveal the distribution of opinion within the field.

**Limitations.** Expert surveys suffer from well-documented biases. Analyses of historical predictions have observed a pattern sometimes called *moving horizon* bias, in which forecasters tend to place AGI roughly 15–25 years in the future regardless of when asked, though this finding is based on a limited dataset (Armstrong and Sotala, 2015). Selection effects may overrepresent researchers who are optimistic about the field's potential (AI Impacts, undated). Expert survey responses also show significant framing effects, with alternative framings producing substantially different results. When researchers were asked about "full automation of labor" rather than "[can] carry out most human professions at least as well as a typical human" (HLMI), median estimates differed by more than 69 years (Grace et al., 2025). Additionally, forecast uncertainty was reflected in the variance between experts in predicting specific scientific milestones. In a 2025 survey, one-quarter of experts gave an 81 percent likelihood of AI solving a Millennium Prize Problem by 2040, while another one-quarter of experts gave only a 30 percent likelihood (Murphy et al., 2025). Most importantly, despite their knowledge of AI systems, experts necessarily lack track records on AGI prediction specifically, making calibration difficult.

## Crowd Forecasting and Prediction Markets

**Method.** Platforms such as Metaculus and Manifold aggregate predictions from large numbers of forecasters, using scoring rules or market mechanisms to incentivize accuracy. These approaches build on research demonstrating that aggregated forecasts from skilled generalists can match or exceed expert predictions on well-specified questions (Tetlock and Gardner, 2015).

**Key findings on AGI timelines.** Metaculus community predictions for AGI have moved earlier, from a median of about 2070 in 2020 to approximately 2033 in early 2026, though these predictions use platform-specific resolution criteria that set a lower bar than many common definitions of AGI. Other prediction markets show similar trends, often with rapid updates to shorter timelines following major model releases (Todd, 2025).

**Strengths.** Aggregation theoretically cancels individual biases. Real-money markets provide financial incentives for accuracy. Prediction market platforms show faster reaction times to new information than shown in annual surveys and have strong calibration track records on well-specified, short-term questions. The Metaculus community prediction, which aggregates all user forecasts, has demonstrated strong calibration across resolved questions (Metaculus, undated).

**Limitations.** Prediction markets struggle with long-horizon questions for which resolution is decades away and bettor's capital is tied up until a question is resolved. Additionally, "reaching AGI" is not a cleanly resolvable question: Platforms must define concrete, verifiable resolution criteria, which necessarily narrows the target relative to broader conceptions of AGI. Because narrower targets are easier to achieve, this structural constraint likely biases prediction market timelines shorter than those from expert surveys that use more-expansive definitions. Participant pools may be systematically biased toward particular technologists with different perspectives than a more representative sample would be. Financial incentives do not uniformly promote accuracy; participants with stakes in AI companies or the broader AI ecosystem may have incentives to distort market prices to influence narratives related to AI timelines, and thin markets on long-horizon questions are particularly susceptible to manipulation by well-capitalized actors. Calibration on short-term questions does not guarantee accuracy on unprecedented transformative events. The Existential Risk Persuasion Tournament found that superforecasters and domain experts continued to disagree substantially about AI risks even after months of structured deliberation, suggesting that short-term forecasting skill may not transfer to long-horizon transformative events (Karger et al., 2025).

## Compute-Centric Models

**Method.** These models estimate AGI timelines by projecting when available computational resources will match some anchor point—typically, estimates of the brain's computational capacity—or by modeling the dynamics of capability growth once AI systems begin contributing to AI research. The most influential framework is Ajeya Cotra's Biological Anchors model developed for Open Philanthropy (now Coefficient

Giving; Cotra, 2020). Davidson (2023) extends the compute-centric approach by explicitly modeling the feedback loop in which AI systems accelerate AI research. His model parameterizes the rate at which AI can substitute for human researchers, hardware cost declines, and training efficiency gains to generate distributions over *takeoff* speeds—the time from human-level AI to substantially superhuman systems. The model's primary contribution is making explicit the assumptions required to generate any takeoff prediction, revealing that disagreements about automation speed largely determine whether one expects gradual transition or rapid capability gains concentrated in a short window.

**Key findings on AGI timelines.** Cotra's 2020 Biological Anchors report estimated a median AGI arrival around 2050, with substantial probability mass in the 2030s and 2040s. However, recent progress has exceeded the assumptions underlying these original estimates. Updates incorporating faster-than-expected algorithmic efficiency gains and compute scaling have shifted probability mass substantially toward earlier dates (Cotra, 2022; Karnofsky, 2021); analyses applying updated trend data to these models yield central estimates around 2030 rather than 2040 (John C, 2025). Davidson's takeoff model finds that, under a wide range of assumptions, the period from "AI that marginally accelerates R&D" to "AI that can fully substitute for human researchers" may be compressed—potentially within a few years rather than decades—because even partial automation creates feedback effects that accelerate subsequent progress.

**Strengths.** Compute models provide mechanistic structure grounded in measurable physical quantities (transistors, FLOPS, dollars). They force explicit statement of assumptions that can be examined and debated. They can incorporate empirical observations about scaling laws, such as the power-law relationships between compute, parameters, and performance documented by Kaplan et al. (2020) and refined by Hoffmann et al. (2022). Davidson's model adds value by connecting timeline questions to takeoff dynamics: Rather than treating AGI arrival as an endpoint, it models what happens next, helping decisionmakers think about the window for response.

**Limitations.** These models rest on highly uncertain parameters: Estimates of brain compute vary by orders of magnitude depending on what is measured and how. Parameter sensitivity is extreme: Modest changes in biological efficiency estimates can shift biological anchors predictions by decades. The mapping from *brain-equivalent compute* to AGI is assumed rather than demonstrated. Davidson's model introduces additional parameters—particularly the rate and ceiling of AI automation of research tasks—that are similarly uncertain and dominate the output. Algorithmic efficiency improvements, which may dominate hardware scaling, are difficult to forecast, especially if driven by unpublished techniques that are proprietary to private companies. Neither model has been subjected to systematic independent replication or adversarial stress-testing.

## Empirical Trend Extrapolation

**Method.** Organizations such as Epoch AI and METR systematically track empirical trends in training compute (the total computational resources used to develop a model), model capabilities, benchmark performance, and algorithmic efficiency (how much capability is achieved per unit of compute), then extrapolate these trends.[1]

**Key findings.** Training compute has grown roughly fourfold to sixfold per year for frontier models since 2010 (Sevilla et al., 2022; Epoch AI, 2025). Scaling laws suggest that model performance improves as a predict-able power law with increases in compute and data (Kaplan et al., 2020). In such relationships, doubling a key input yields a consistent fractional improvement in output. Benchmark performance has improved rapidly, with many tests that were designed to challenge models for years becoming saturated (models reaching near-ceiling scores) within months. Massive Multitask Language Understanding (MMLU) accuracy rose from approximately 44 percent in 2020 (GPT-3) to over 86 percent by multiple frontier models by 2024, approaching human expert–level performance of approximately 90 percent (Wang et al., 2024; Hendrycks et al., 2021). Performance on Humanity's Last Exam improved from initial results of roughly 8 percent (OpenAI's o1) and 11 percent (Google's Gemini 2.5 Flash) to 35 percent (GPT-5.2) and 37 percent (Gemini 3 Pro) (Center for AI Safety, Scale AI, and HLE Contributors Consortium, 2026). Empirical work by METR in 2025 introduced a new metric: the "50% Task Completion Time Horizon," defined as the length of tasks—measured by how long they take human professionals—that generalist frontier model agents can complete autonomously with a 50 percent success rate (Kwa et al., 2025). From 2019 to 2025, this horizon doubled approximately every seven months, though more recent data suggest that the trend may be accelerating (Kwa et al., 2025). If the longer-run trend holds, agents could execute weeklong subprojects by the late 2020s; if the recent acceleration persists, this threshold could be reached as early as late 2026 or early 2027, potentially unlocking the automated R&D feedback loop (Lifland, Jurkovic, and FutureSearch, 2025; Kwa et al., 2025).

---

[1] Epoch AI has emerged as one of the most prolific contributors to the AGI forecasting ecosystem. In addition to the compute trends database discussed here, the organization has developed the "Direct Approach," a compute-centric timeline model that bypasses biological reference points (Barnett and Besiroglu, 2023), and the GATE model, which combines AI capability forecasting with macroeconomic automation modeling (Erdil et al., 2025).

**Strengths.** These approaches are grounded in observable data rather than theoretical estimates. They can identify empirical regularities and provide evidence for or against specific hypotheses about the drivers of progress. Similar extrapolation methods have historically performed reasonably well in adjacent domains (e.g., Moore's Law), and AI-specific scaling laws have shown surprising consistency across multiple-order-of-magnitude increases in compute. Additionally, multiple independent trend lines (compute growth, benchmark saturation rates, task-horizon expansion) can corroborate or challenge one another, increasing confidence when trends converge and flagging uncertainty when they diverge.

**Limitations.** Trend extrapolation assumes continuity; it cannot predict discontinuities (breakthroughs or plateaus) that may fundamentally alter trajectories. The relationship between metrics that are measurable (loss—the error rate models are trained to minimize—and benchmark scores) and the target of interest (general intelligence) is unclear. Goodhart's Law poses a significant risk: When a measure becomes a target, it ceases to be a good measure.[2] Research on AI benchmarks has documented systemic issues, including data contamination (test questions appearing in training data), construct validity (whether benchmarks measure what they claim to measure), and spurious correlations that inflate apparent performance (Raji et al., 2021; Eriksson et al., 2025).

## The Market Weight Signal

The preceding methods attempt to directly estimate AGI timelines. A distinct and indirect signal comes from the investment behavior of major technology firms, which constitutes an implicit forecast. Companies such as Microsoft, Google, Amazon, and Meta are investing more than $300 billion annually in AI infrastructure based on expectations of transformative capabilities (Subin, 2025). OpenAI has explicitly stated that its mission is "to ensure that artificial general intelligence . . . benefits all of humanity" (OpenAI, undated), while Anthropic has predicted that "powerful AI systems will emerge in late 2026 or early 2027" (Anthropic, 2025).

---

[2]  Named after British economist Charles Goodhart, who observed that statistical regularities tend to collapse once pressure is placed on them for policy purposes. In the AI context, if a benchmark is used to evaluate progress, developers may optimize specifically for that benchmark rather than the underlying capability it was meant to measure.

**Interpretation.** Previous AI investment cycles never approached this scale. The 1980s expert system boom—the largest prior wave—produced a flurry of start-ups and government initiatives, but even its most ambitious single project, the Cyc knowledge base, received roughly $500 million over a decade (Haigh, 2024). Current annual AI infrastructure spending by major technology firms exceeds that figure by over two orders of magnitude. This represents "skin in the game" that differentiates current AI development from previous hype cycles.

However, this signal has important limitations. The firms making the largest investments are often the same ones making the most aggressive timeline claims, creating a circularity in which corporate forecasts and capital allocation reinforce each other rather than serve as independent confirmation. Competitive dynamics may also drive investment: Firms may spend not because they have independently concluded that transformative AI is imminent but because falling behind a competitor that achieves it first would be catastrophic. Additionally, the concentration of capital among a small number of hyperscalers with few comparably attractive alternative investments means that the sheer scale of spending may reflect market structure as much as informed conviction about timelines.

Across these methods, the directional convergence is notable: An independent European policy assessment reviewing the same evidence concluded that AGI could plausibly emerge between 2030 and 2040 or earlier (Negele et al., 2025).

## Interpreting the Forecasting Ecosystem: Three Archetypes

The preceding sections have surveyed multiple forecasting inputs: expert surveys, prediction markets, superforecaster judgments, scaling analyses, and revealed preferences through capital allocation. These sources offer varying timelines and levels of confidence, which can appear contradictory or difficult to synthesize. Rather than treating each forecast as an independent data point to be averaged, decisionmakers may find it more useful to understand the underlying models of technological change that generate different predictions. Expert views on AGI timelines can be organized into three archetypes, each representing a coherent position on how AI progress unfolds:

**Scaling maximalists** expect continued scaling of current architectures and AI-assisted research to yield AGI as soon as the late 2020s or early 2030s. Scaling may benefit from incremental algorithmic improvements, but such improvements are incidental compared with the effects of adding more compute. This view drives capital expenditure and security concerns. AI 2027 (Kokotajlo et al., 2025), a scenario exercise tracing a specific trajectory from current capabilities through rapid capability gains and AI-automated R&D, offers the most detailed public articulation of this position (see also Aschenbrenner, 2024).

**Paradigm shift advocates** see AGI as plausible in the next few decades but expect that existing deep learning foundations, while sound, are insufficient on their own. This camp spans a wide range: Some anticipate that targeted algorithmic innovations (improved architectures, better integration of symbolic reasoning with neural networks) will bridge the gap, placing AGI in the 2030s to 2040s. Others are more skeptical, arguing that fundamental limitations, such as data scarcity, brittleness, and lack of causal understanding, will require more-radical departures from current methods, potentially pushing timelines into the 2050s or later. What unites this group is the belief that the path to AGI runs through deep learning but requires significant work beyond scaling. Whether that work is evolutionary or revolutionary is the internal disagreement. Yann LeCun, Meta's chief AI scientist and Turing Award laureate, offers one of the most detailed articulations of this position. LeCun argues that autoregressive LLMs are fundamentally incapable of achieving human-level intelligence because they lack world models, sensory grounding, and the capacity for planning. His proposed alternative—the Joint Embedding Predictive Architecture—represents a nongenerative framework designed to learn abstract representations of the world through observation rather than token prediction (LeCun, 2022). LeCun has stated that reaching human-level AI "will take several years if not a decade" while noting that the distribution "has a long tail" and could take much longer (LeCun, 2024).

**Economic skeptics** occupy a different position from the preceding two. Their skepticism is not primarily about whether AGI-level capabilities will emerge but about whether such capabilities will translate rapidly into economic transformation. One can hold short technology timelines while still expecting slow diffusion—making this a distinct dimension of disagreement rather than a point on the same spectrum. Slow transformation could result from deployment bottlenecks (manufacturing, robotics, regulatory barriers), organizational adjustment costs (the productivity J-curve), or political economy constraints (labor opposition, public backlash) (Narayanan and Kapoor, 2025). Economic skeptics may hold a range of views on technical timelines but predict economic transformation in the 2040s or later.

These categories are ideal types; individual experts may hold hybrid views or shift positions as evidence accumulates. Dario Amodei, chief executive officer of Anthropic, exemplifies the scaling maximalist position: He has publicly stated that powerful AI "could come as early as 2026" and described the mechanism as a self-reinforcing loop in which AI systems that are good at coding and research produce the next generation of models (Amodei, 2024). At Davos in January 2026 he reaffirmed this timeline, predicting AI models performing at Nobel laureate level across multiple fields in one to two years (Amodei, 2026). Gary Marcus sits at the more skeptical end of paradigm shift advocates, combining concerns about current architectures with deep skepticism about deep learning's ultimate ceiling (Marcus, 2022). For decisionmakers, the value of this typology lies not in picking a winner but in stress-testing strategy against multiple scenarios. A policy calibrated to scaling maximalist timelines may prove premature if paradigm shift advocates are correct, but waiting for certainty may leave society unprepared if the maximalists prove right. Under the most compressed timelines, moreover, the window for adjusting policy and institutional response capabilities based on accumulating evidence may itself be narrow, which argues for robust preparation across scenarios rather than sequential adaptation.

## Chapter 3 References

# Understanding Disagreement: The Cruxes

Experts disagree about AGI timelines because they hold different views on specific empirical and conceptual questions. Identifying these specific sources of disagreement helps decisionmakers understand what evidence would cause experts to update forecasts and where to focus research. We call these *cruxes*: specific empirical questions for which different answers generate different timeline estimates. We have identified three cruxes: capability, diffusion, and takeoff. Complementary work has proposed structured dimensions—such as centralization of ownership, governance primacy, and takeoff speed—for classifying and comparing potential AGI futures once timelines are assumed (Chessen and Chowdhury, 2025).

## The Capability Crux: Scaling, Architecture, and the Limits of Current Approaches

**The question:** Will scaling existing deep learning architectures with more compute and data be sufficient to reach AGI, or will fundamentally new approaches be required?

This is arguably the central disagreement in AGI forecasting. It encompasses two closely related questions that are often treated separately but share a common structure: whether scaling laws will continue to deliver capability gains at frontier compute levels and whether the capabilities that scaling delivers will ultimately be *sufficient* for general intelligence. Both questions concern the distance between where we are and where AGI is: The first asks whether the current trajectory continues, and the second asks whether the trajectory is pointed at the right destination.

**The "continuation and sufficiency" position:** Scaling laws have held over several orders of magnitude (Kaplan et al., 2020; Hoffmann et al., 2022). There is no principled reason to expect they will fail at any particular scale. Synthetic data and improved training techniques can address data scarcity. Current architectures, properly scaled with sufficient data and appropriate scaffolding (tool use, retrieval, chain-of-thought reasoning), are learning something genuinely general about intelligence—as suggested by the breadth of capabilities that emerge at scale (Wei et al., 2022). Under this view, AGI does not require a fundamentally new paradigm; it requires executing the current one at sufficient scale, with incremental architectural improvements along the way.

**The "plateau and insufficiency" position:** This camp spans a wide range of views united by the belief that current approaches face limits that scaling alone cannot overcome. At the nearer end, some researchers expect that targeted innovations—improved architectures, better integration of symbolic reasoning with neural networks, or new training paradigms, such as LeCun's proposed Joint Embedding Predictive Architecture (LeCun, 2022)—will bridge the gap. At the more skeptical end, critics argue that current systems lack fundamental capacities, such as world models and causal reasoning (LeCun, 2022), and that fluent language production does not entail genuine understanding (Bender et al., 2021). Bender and colleagues coined the term "stochastic parrots" to describe systems that produce fluent language by pattern-matching over training data without underlying comprehension. This framing remains contested—proponents of the sufficiency position argue that it underestimates what pattern-matching at scale can achieve—but it captures a real concern about the gap between benchmark performance and robust general capability.

A key variant of this position is that scaling may be insufficient in a particularly strong sense: Not only might current architectures lack the capacity to reach AGI, but they may also lack the capacity to *invent the architectures that would*. Under this view, fundamental breakthroughs in such areas as neurosymbolic integration or world modeling require the kind of creative conceptual leaps that current systems cannot reliably produce, making human-driven paradigm shifts a prerequisite.

High-quality training data are also finite: Estimates suggest that language models will have consumed the available stock of high-quality human-generated public text data by 2026–2032 (Villalobos et al., 2024), with the lower bound already upon us. Whether synthetic data and other techniques can fully substitute for this resource is an open question.

**The evidence:** Scaling laws have been remarkably consistent within the deep learning paradigm. Frontier systems demonstrate competent performance in coding, mathematics, medicine, and general knowledge across diverse domains, representing a significant departure from the narrow AI systems of the previous decade. At the same time, models show failures on out-of-distribution tasks and inconsistent multistep reasoning that raise questions about the depth of their capabilities (Mirzadeh et al., 2024), although the severity of these failures and whether they represent fundamental limits or problems that further scaling will address are contested. Some research has argued that apparent "emergent abilities" may be artifacts of metric choice: When researchers use log-probability scoring rather than discrete accuracy thresholds, the sharp transitions attributed to emergence disappear in favor of smooth, predictable improvements (Schaeffer, Miranda, and Koyejo, 2023). This finding cuts in multiple directions: It challenges claims of sudden capability jumps but is also consistent with steady, predictable progress driven by scale.

**What would resolve it:** The most informative evidence will come from observing capability trajectories at the \$10 billion or greater training run scale. The key question is not simply whether loss curves continue to improve—proponents on both sides expect continued benchmark gains—but whether improvements generalize to genuinely novel tasks and domains. Specifically, do frontier models at the next compute threshold show robust performance on out-of-distribution reasoning, on tasks requiring causal inference, and on multistep problems with long time horizons? If so, the continuation and sufficiency position gains significant support. If performance on such tasks plateaus or remains brittle despite massive compute increases, the case for architectural insufficiency strengthens. Development of evaluation methods that can distinguish genuine generalization from sophisticated interpolation would be particularly valuable, though designing such evaluations is itself a significant, unsolved challenge.

## The Diffusion Crux: Speed of Transformation

**The question:** If AGI-level capabilities are achieved, how quickly will they transform the economy and society?

**The "rapid diffusion" position:** Current AI adoption is already historically fast: ChatGPT reached 100 million users within two months of launch, making it one of the most rapidly adopted software products ever (Hu, 2023). Proponents of rapid diffusion argue that this speed, while striking, is not the core concern—rather, it is what happens after highly capable or superintelligent systems emerge. Such systems might overcome normal adoption barriers by flexibly adapting to organizational contexts, solving their own deployment bottlenecks, and learning on the job in ways that existing AI cannot (Aschenbrenner, 2024). Competitive pressure (commercial and geopolitical) would intensify the urgency to adopt. Davidson's take-off model formalizes this intuition: Even partial automation of cognitive labor creates feedback effects that accelerate subsequent adoption, potentially compressing diffusion timelines relative to historical precedent (Davidson, 2023). Under this view, historical precedents of slow-diffusing, general-purpose technologies may not apply to systems capable of actively accelerating their own integration.

**The "slow diffusion" position:** Regulatory approval, liability concerns, organizational change management, and infrastructure requirements will delay adoption regardless of capability level. Physical-world transformation requires robotics, manufacturing, and infrastructure that cannot be instantiated overnight. Public resistance, labor opposition, and political backlash may further constrain deployment—particularly in domains in which AI threatens employment, raises safety concerns, or challenges existing institutions. Historical transformation from general-purpose technologies took decades (David, 1990; Narayanan and Kapoor, 2025), and Ding (2024) argues that, in past technological revolutions, what determined shifts in great-power competition was not which nation first pioneered the technology but which one built the institutional capacity to diffuse it broadly across the economy (see also Mazarr, 2026). Under this view, software adoption speed is a misleading comparison: The relevant benchmark is deep economic integration, not user sign-ups. Organized opposition to AI deployment, including movements calling for pauses or moratoriums on frontier development, regulatory intervention, and institutional resistance, could further constrain the pace of transformation independently of technical capability.

**The evidence:** Current evidence is mixed. The rapid initial adoption noted above is historically unprecedented, but depth of integration tells a more complex story. On task-level productivity, results are strikingly context dependent. A field experiment at Boston Consulting Group found that consultants using GPT-4 completed significantly more tasks at higher quality when working within the AI's capability range but performed substantially worse on tasks outside that range, a pattern the researchers term the "jagged technological frontier" (Dell'Acqua et al., 2023). More recently, METR's randomized controlled trial found that experienced open-source developers using frontier AI coding tools were actually 19 percent slower than without AI assistance, even though the developers believed they had been sped up by 20 percent (Becker et al., 2025). These task-level findings have not appeared in aggregate economic statistics either: Research tracking labor market outcomes in Denmark found minimal macroeconomic impact through early 2024 (Humlum and Vestergaard, 2025), consistent with the "productivity J-curve" hypothesis, under which general-purpose technologies require extended periods of complementary investment before aggregate effects materialize (Brynjolfsson, Rock, and Syverson, 2021). However, as proponents of rapid diffusion note, this lag is precisely what historical precedent would predict even for technologies that ultimately prove transformative—electricity did not produce measurable productivity gains for decades after its introduction (David, 1990)—and current adoption speed far outpaces those historical comparators. Proponents of rapid diffusion further argue that current productivity measurements are uninformative about post-AGI dynamics, when systems capable of overcoming their own deployment bottlenecks could render historical diffusion patterns obsolete.

**What would resolve it:** Unlike the other two cruxes, this one may be difficult to resolve incrementally. Current adoption and productivity metrics cut both ways: Unprecedented uptake speed suggests that diffusion barriers may be lower than historical precedent implies, while the absence of aggregate economic effects and the mixed task-level results are consistent with either genuinely slow integration or the early stages of a productivity J-curve. The disconnect between perceived and measured productivity gains—as in the METR finding—adds a further complication: Subjective reports of AI usefulness may not reliably track actual economic impact. Resolution may require observing how systems far more capable than today's interact with deployment barriers or developing theoretical frameworks for what capability thresholds would suffice to overcome organizational and regulatory friction.

# The Takeoff Crux: Speed of Capability Gain

**The question:** Once AI systems reach human-level performance on key cognitive tasks, how quickly will capabilities improve further—and will this improvement be observable in time to respond?

**Defining "takeoff":** Different researchers operationalize this question differently. Bostrom (2014) defines takeoff in terms of *capability trajectory*: the clock time from human-level AI to superintelligent AI, with "slow" being decades to centuries, "moderate" being months to years, and "fast" being minutes to days. Christiano (2018) defines takeoff in terms of economic transformation: Slow takeoff means the economy experiences a four-year doubling of output before experiencing a one-year doubling—still extremely rapid by historical standards, but continuous rather than discontinuous. These definitions can diverge if rapid capability gains occur inside laboratories before deployment. The analysis below focuses primarily on capability trajectory, treating economic transformation under the separate diffusion crux.

**The "gradual takeoff" position:** Capability improvements will be continuous and observable (Christiano, 2018). Before AI systems can dramatically accelerate AI development, they will first *modestly* accelerate it; before that, they will *slightly* accelerate it. Each improvement will be deployed and generate observable effects before the next major advance. Under this view, the path from human-level AI to substantially superhuman AI would unfold over years, providing time to observe progress, develop safety measures, and adapt governance. Weaker AI systems will already radically transform the world before full AGI arrives, offering warning signs (Davidson, 2023). Notably, even aggressive timeline forecasts—such as AI 2027—may be consistent with gradual takeoff: Reaching AGI soon does not imply that the subsequent path to substantially superhuman capabilities will be discontinuous (Kokotajlo et al., 2025).

**The "rapid takeoff" position:** If AI systems become capable of substantially automating AI research, a feedback loop could compress years of capability progress into months or weeks (Good, 1966; Bostrom, 2014). Crucially, this acceleration could occur *inside AI laboratories* before significant economic deployment—an intelligence explosion that outpaces external observation. The transition from human-level to substantially superhuman capability might happen faster than governance systems can track, even if diffusion into the broader economy remains slow. Under Bostrom's taxonomy, this corresponds to "fast" or "moderate" takeoff; under Yudkowsky's framing, this is the "FOOM" scenario, in which a single project's capabilities rapidly outpace the rest of the world (Bostrom, 2014; Yudkowsky, 2008).

**The "persistent bottleneck" position:** A third view holds that fundamental constraints will limit the pace of capability improvement regardless of AI's role in research. Under this view, a decade or more might separate AI systems capable of fully automating R&D from genuinely superintelligent systems—those whose cognitive capabilities substantially exceed human performance across virtually all domains (Bostrom, 2014)—because algorithmic progress faces diminishing returns or because key bottlenecks lie in hardware, data, or theoretical insights that AI cannot easily accelerate. This position differs from gradual takeoff: Whereas Christiano (2018) expects dramatic capability acceleration that nonetheless unfolds continuously, the bottleneck view expects constraints to fundamentally limit how fast capabilities can improve. Critics argue that this position underestimates both the potential speedup from AI-driven research and the ability of highly capable systems to route around apparent bottlenecks.

**The evidence:** Current AI systems are beginning to assist with AI research—code generation, experiment design, literature review—though the extent to which they have displaced human researchers at the frontier remains unclear. Empirical data on AI automation of AI research remains limited, though emerging work decomposes AI R&D into subtasks and elicits expert estimates of automation potential (Owen, 2024). Research on benchmarks in 2024 found that AI agents could outperform human experts on short-duration research engineering tasks, though humans still displayed better returns on extended time budgets (Wijk et al., 2024).

Davidson (2023) provides a formal framework for analyzing these scenarios. His compute-centric model finds that, under assumptions in which AI eventually automates most cognitive labor involved in R&D, even modest initial automation rates produce substantial acceleration effects. The model suggests that the gradual and rapid positions may be less distinct than they appear in practice: Under many parameterizations, the world is already substantially transformed by the time systems reach full human-level capability.

**The relationship to diffusion:** The takeoff and diffusion cruxes are related but distinct. Takeoff concerns how quickly *capabilities improve*; diffusion concerns how quickly *achieved capabilities transform the economy*. Rapid capability takeoff with slow diffusion would mean fast progress inside laboratories without immediate economic effects—a scenario some consider particularly concerning because external observers would have limited visibility into capability gains until deployment decisions are made. Conversely, gradual capability takeoff with rapid diffusion would mean that modest improvements are quickly adopted, producing substantial economic transformation even without dramatic capability jumps. The coupling between these two processes is contested: Gradual takeoff proponents generally expect tight coupling, whereas rapid takeoff proponents argue that the processes could substantially decouple.

**What would resolve it:** Metrics tracking AI contribution to AI research would provide the most direct evidence. If the fraction of frontier AI research automated by AI systems increases rapidly, scenarios involving significant capability acceleration become more plausible—though acceleration of AI R&D inputs does not guarantee acceleration of AI R&D outputs, since increased automation may simply be the necessary next step to maintain the current pace of progress rather than a sign of imminent takeoff. Current indicators suggest that direct automation remains limited but is increasing: Interviews with AI researchers point to partial automation of time-consuming engineering work (Owen, 2024), and frontier AI "time horizons" on research-like tasks have been doubling roughly every seven months since 2019 (Kwa et al., 2025). The latter metric—how long a task an AI system can complete autonomously—may prove more informative than traditional benchmarks for tracking progress toward the capability thresholds that would enable rapid takeoff. However, this evidence is largely self-reported by researchers at frontier laboratories with incentives to emphasize AI's growing contribution. The most informative data on AI's role in AI research—internal productivity metrics, commit logs, research pipeline analysis—is proprietary to the labs. Establishing mechanisms for sharing such data with regulators or independent evaluation bodies, potentially through a trusted consortium, would substantially improve the forecasting community's ability to track this critical variable.

## Summary: How Cruxes Generate Positions

In the previous chapter, we introduced three archetypes that characterize expert views on AGI timelines. The cruxes analyzed in this chapter reveal what drives these positions apart: Each archetype reflects a distinct combination of views on capability trajectories, diffusion speed, and takeoff dynamics to match the merged crux, as summarized in Table 4.1.

These positions are idealized points along a spectrum, not rigid categories. Individual experts often hold hybrid views: A researcher may expect continued scaling gains while anticipating significant architectural changes between scale-ups or may hold short technical timelines while remaining uncertain about diffusion speed. The value of this framework lies not in selecting which row is correct but in identifying what to watch. Each of these disagreements suggests specific observables: whether frontier models generalize robustly at the next compute threshold (capability), how quickly AI tools achieve deep integration into economic activity (diffusion), and what fraction of AI research is automated by AI systems (takeoff). As evidence accumulates on these questions, positions should update accordingly, and decisionmakers can anticipate which strategic implications would shift with which observations. However, under the most compressed timelines, this sequential observe-and-update model may be unrealistic: If AGI-level capabilities emerge in the late 2020s,

**TABLE 4.1**

**Crux Positions Underlying Each Archetype**

| Position | Key Assumptions | Implied Timeline | Strategy and Policy Implications |
|---|---|---|---|
| Scaling Maximalist | Current architecture is sufficient with incremental improvements; rapid diffusion likely; AI-driven R&D acceleration compresses takeoff | Late 2020s to early 2030s | Urgent safety investment; prepare for rapid societal transformation |
| Paradigm Shift Advocate (nearer end) | Scaling continues, but targeted architectural innovations are needed; diffusion pace uncertain; takeoff gradual | Mid-2030s to 2040s | Build capacity steadily; monitor key indicators |
| Paradigm Shift Advocate (skeptical end) | Deep learning hits fundamental limits; new paradigm required; takeoff question is premature until capability gap is closed | 2050s or later | Focus on narrow AI governance; avoid overpreparation costs; invest in diverse AI approaches at the national level |
| Economic Skeptic | Technical progress may be fast; economic transformation is slow because of deployment barriers; takeoff in labs may decouple from diffusion into economy | Technical: 2030s; Economic: 2050s or later | Distinguish capability from impact; reactive labor policy |

the window for incremental evidence-gathering is narrow, and the most consequential decisions may need to be made before these pivotal questions are resolved. This asymmetry argues for preparing across scenarios now rather than waiting for clarity that may arrive too late to act on. These decisionmakers span different contexts: government agencies setting compute governance policies, AI labs allocating research priorities, funding organizations directing grant portfolios, and companies planning workforce transitions. Each will need to track different observables depending on their strategic position, but all can benefit from understanding which empirical developments would shift the landscape.

## Chapter 4 References

# The Validation Challenge: Limitations of Current Methods

The forecasting approaches described in the previous chapters share a fundamental problem: They lack the validation infrastructure needed to assess their reliability. We cannot evaluate forecasters' track records on AGI prediction because AGI has not arrived. The benchmarks used to track progress are saturating and gameable. Even when benchmarks compare AI with human performance, the human baselines often lack the rigor and transparency needed to support reliable comparisons (Wei et al., 2025). The influential models that shape expectations have not been independently stress-tested. This validation challenge does not invalidate all forecasting; some methods and signals are more reliable than others. But it should prompt caution and motivate investment—by research funders, government agencies, and frontier laboratories—in better measurement infrastructure.

It should also be acknowledged that the positions synthesized in this report are not all mutually compatible. The empirical cruxes identified in Chapter 4—about scaling, architecture, and takeoff dynamics—often point in contradictory directions, and the range of timelines they imply span months to decades. Under the shortest credible timelines, many of the research investments and institutional preparations discussed in later chapters would not mature in time to be useful. This does not diminish their value as a portfolio, but decisionmakers should understand that no single reading of the evidence validates the entire agenda.

## The Benchmark Problem

Public benchmarks are the primary instrument for tracking AI progress, yet they are failing as reliable measures. The core problem is a variant of Goodhart's Law: When a measure becomes an optimization target, it ceases to be a good measure of the underlying construct it was designed to assess (Strathern, 1997). This dynamic manifests in three interrelated ways.

**Saturation.** Benchmarks that were designed to remain challenging for years can become saturated within months of release, reducing their ability to discriminate between systems or track further progress. MMLU, introduced in 2020, was intended as a difficult test of expert-level knowledge across 57 subjects. At that time, GPT-3 scored only about 43 percent, compared with human expert performance around 90 percent (Hendrycks et al., 2021). By 2023, GPT-4 achieved 86.4 percent on MMLU, close to human experts (OpenAI et al., 2024). Since then, many frontier models have clustered in the high 80s on MMLU, with only a few percentage points separating them, making it increasingly difficult for the benchmark to distinguish among state-of-the-art systems (Wang et al., 2024). Whether this reflects genuine capability convergence or optimization toward a common target is unclear.

**Gaming and contamination.** As benchmarks become explicit optimization targets, high scores may reflect training data contamination, benchmark-specific heuristics, or focused optimization rather than general capability. Research has found that, in the MMLU benchmark, GPT-4 could correctly guess missing answer options 57 percent of the time, raising concerns about potential exposure to test data during train-ing (Deng et al., 2024). A comprehensive survey of benchmark data contamination documented that significant overlap between training and evaluation data is pervasive across popular LLM benchmarks, leading to inflated performance scores (Xu et al., 2024). A model scoring 95 percent on a reasoning test may not be able to reason; it may have learned to recognize and solve that specific test format.

**Construct validity.** Even absent contamination, it is unclear what benchmarks actually measure. Do high scores on coding challenges indicate genuine programming ability or pattern-matching to training data? The Abstraction and Reasoning Corpus (ARC-AGI) was designed by François Chollet (2019) specifically to test general fluid intelligence through novel reasoning tasks that would resist memorization and statistical pattern-matching. ARC-AGI-2 is an expanded, more challenging successor that preserves the same task format while reducing susceptibility to brute-force strategies and improving discrimination at higher difficulty, calibrated using first-party human data (Chollet et al., 2025). Yet even this benchmark shows rapid progress, with Google's Gemini 3 Deep Think scoring roughly 85 percent and Anthropic's Claude Opus 4.6 scoring roughly 70 percent.[1] Whether these scores represent genuine advances in abstract reasoning or increasingly sophisticated approaches to the benchmark's specific structure remains unclear, illustrating the fundamental challenge of creating durable measures of general intelligence.

## Immaturity of the Forecasting Ecosystem

Unlike fields such as weather forecasting or macroeconomic modeling, in which competing research groups, national agencies, and decades of iterative refinement have produced mature modeling ecosystems, compute-centric AGI forecasting remains a young field with few independent models and modelers (Cotra, 2020; Karnofsky, 2021). The field more closely resembles early-stage climate projection or pandemic preparedness modeling: domains in which the stakes are high, the underlying dynamics are poorly constrained, and the relevant events are too rare to provide regular calibration feedback.

**Parameter uncertainty.** These models require estimates of quantities, such as the computational capacity of the human brain or algorithmic efficiency relative to biological neurons. These estimates span orders of magnitude depending on what is measured and how (Carlsmith, 2020). The models acknowledge this uncertainty through wide probability distributions, but the outputs remain highly sensitive to input assumptions. As critics have noted, uncertainty in key parameters—such as how much of the brain's computational activity is relevant to general intelligence—may dominate the entire timeline conversation, and people with different initial beliefs about these parameters may be largely unmoved by the model's calculations (Lin, 2022).

**Insufficient independent validation.** Influential forecasting models are typically produced by single teams or individuals, and the forecasting ecosystem remains small. Independent researchers have not systematically attempted to reproduce results, identify coding errors, or test sensitivity to alternative assumptions—standard practice in other scientific domains with high-stakes predictions. Nor has there been substantial investment in developing alternative modeling approaches or in extending and refining existing frameworks. Although the Biological Anchors report received thoughtful critiques examining its central assumptions (Lin, 2022; Yudkowsky, 2021), the field lacks the competing research programs, derivative implementations, methodological diversity, and accumulated institutional knowledge that strengthen predictions in more mature forecasting domains.

---

[1]  This is a snapshot of scores as of February 20, 2026.

## Institutional Independence Concerns

Recent episodes illustrate the consequences of limited transparency and independence. In the FrontierMath controversy, for example, OpenAI funded the benchmark's development and had access to most problems and solutions—facts withheld from contributing mathematicians and disclosed only on the day OpenAI announced its o3 model's performance (Wiggers, 2025a; Besiroglu and Sevilla, 2025). Similarly, Meta submitted a specially tuned "experimental" version of its Llama 4 Maverick model to the LM Arena leaderboard that was optimized for conversational style and differed substantially from the publicly released model, leading to accusations of benchmark manipulation (Wiggers, 2025b). Such arrangements raise questions about benchmark integrity when labs can selectively tune or fund evaluations. More broadly, research has documented that model creators frequently report benchmark scores using nonstandard prompting techniques, making controlled comparisons difficult, and that third-party evaluations often produce substantially lower scores than those reported by developers (Mai and Liang, 2024). The field needs clearer disclosure of funding and institutional ties, public documentation of data sources and cleaning procedures, and open-sourcing of code and datasets where feasible.

## The Calibration Problem

Although many forecasters and forecasting platforms have begun to build up enough of a track record for independent evaluation, several fundamental issues limit our ability to assess calibration on AGI prediction specifically.

**No resolved forecasts.** AGI is a one-time event that has not occurred. We cannot evaluate forecasters based on their accuracy in predicting it. Historical predictions about AI timelines were largely made informally, with fuzzy targets and no systematic tracking. Analysis of 95 AI timeline predictions found that expert forecasts were "indistinguishable from non-expert predictions and past failed predictions" (Armstrong and Sotala, 2015).

**Limited transferability.** Forecasters may be well-calibrated on short-term, well-specified questions (e.g., "Will an AI system score above 90 percent on MMLU by 2025?") while poorly calibrated on long-term, fuzzy questions (e.g., "When will systems achieve human-level general intelligence?"). Calibration in one domain does not guarantee calibration in another.

**Expert limitations.** A further issue is the mismatch between domain expertise and forecasting expertise. AI researchers have domain expertise but not forecasting expertise. Superforecasters have forecasting expertise but not domain expertise. It is unclear which expertise matters more for AGI prediction, and there is no empirical basis for weighting them (Tetlock and Gardner, 2015).

## Reflexivity: Forecasts Shape What They Predict

An often-overlooked feature of forecasting is that forecasts and strategy interact. Forecasts help shape the trajectory they aim to predict. Optimistic forecasts drive funding, attract talent, and signal regulatory urgency, which may accelerate development. Pessimistic forecasts may reduce investment or induce complacency, which may slow it.

This reflexivity is not inherently problematic; it can serve as a coordination mechanism. Moore's Law, for instance, began as an empirical observation but evolved into a shared industry roadmap. As Gordon Moore himself noted, "Once something like this gets established, it becomes more or less a self-fulfilling prophecy . . . . Everyone in the industry recognizes that if you don't stay on essentially that curve they will fall behind" (Schaller, 2004, p. 390, quoting Moore, 1996). The Semiconductor Industry Association's technology road-maps used Moore's Law as a baseline, enabling firms, investors, and governments to coordinate long-term planning and R&D investment (Semiconductor Industry Association, 1994). A forecast that becomes a coordinating target can be genuinely useful.

Decisionmakers should therefore understand that AGI forecasts are not neutral observations of an external reality; they are inputs into a complex system that responds to expectations. This creates both opportunities and risks. Shared timelines can help coordinate safety research, infrastructure investment, and regulatory preparation. But forecasting communities and labs also share social networks and incentives, which can lead to correlated errors. Optimistic or pessimistic narratives may become self-reinforcing or provoke regulatory responses that alter the trajectory in unintended ways. Decisionmakers should account for this reflexivity when interpreting and acting on forecasts.

## Foundational Objections

The preceding sections documented methodological limitations—benchmark validity, calibration, institutional independence—that constrain forecasting reliability. Some AI researchers raise a more fundamental objection: that forecasting "AGI arrival" may be asking the wrong question entirely.

This critique holds that AGI is not a coherent threshold to predict. AI capabilities emerge unevenly across domains—what Dell'Acqua et al. (2023) call a "jagged technological frontier"—in which a system might demonstrate expert-level coding while failing elementary physical reasoning or produce fluent prose while struggling with basic arithmetic. If capability profiles remain jagged rather than converging toward uniform competence, then asking "when will AGI arrive?" faces a measurement problem: There is no single underlying dimension to track and no agreement on how to aggregate across domains. However, this objection has limits. If AGI is defined as performing *all* tasks at human expert level, then the definition remains coherent even in the presence of jaggedness; the jagged frontier simply has not yet exceeded the human baseline everywhere. The practical forecasting challenge is that jaggedness makes intermediate progress hard to interpret: A system that exceeds human performance in 80 percent of domains may be either very close to AGI or very far from it, depending on what remains in the final 20 percent.

A related concern is that current disagreements may reflect something deeper than empirical uncertainty. The forecasting disputes documented in this report—about scaling, architecture, and the nature of intelligence itself—bear hallmarks of what Thomas Kuhn called a pre-paradigmatic state: a field in which practitioners lack shared standards for what counts as evidence, what the right questions are, and how to interpret observations (Kuhn, 2012). If so, these disputes may not be fully resolvable through additional data alone; they may await a conceptual breakthrough that reframes the question entirely.

These objections have merit but do not eliminate the value of forecasting. Even if *AGI* is imprecise and the field pre-paradigmatic, the underlying questions—about capability trajectories, automation potential, and the pace of progress—remain decision-relevant. The appropriate response is not to abandon forecasting but to be explicit about what is being predicted and to track multiple capability dimensions rather than a single threshold. If the field is indeed in a Kuhnian crisis, the value of current forecasting may lie less in convergence toward a shared answer than in clarifying the assumptions and commitments that divide it, surfacing what would otherwise remain implicit, discriminating among possible futures by emphasizing what is most operationally significant, and structuring preparation for a range of outcomes.

## What Can We Rely On?

Despite these limitations, some signals are likely more reliable than others. The signals reviewed here, and summarized in Table 5.1, were selected because they are commonly cited in strategic discussions and have relatively clear empirical grounding. They are primarily *input* signals (compute, capital, scaling trends) or *aggregate judgments* (expert surveys, prediction markets). An alternative approach would track *capability milestones*—specific, demonstrated abilities that might mark waypoints toward AGI, such as those proposed in Google DeepMind's "Levels of AGI" framework (Morris et al., 2024) or Li and colleagues on dangerous capability benchmarks (Li et al., 2024). We address capability-based indicators separately in Chapter 6, as they raise distinct measurement challenges discussed at the beginning of this chapter.

The implication is not that forecasting is worthless. The directional signal of shifted timelines and increased investment is meaningful.[2] But decisionmakers should hold specific year predictions loosely and focus on building capacity for a range of scenarios.

**TABLE 5.1**
**Signals of Timeline Trajectory**

| Signal | Signal Robustness | What It Shows | Key Limitation |
|---|---|---|---|
| Compute trends | High | Physical resources available for training | Compute is necessary but not sufficient for capability |
| Capital investment | High | Industry belief in near-term progress | Markets can be irrational; bubbles or underinvestment can occur |
| Shifted or compressed timelines | High | Direction of movement in AGI timelines across expert surveys, prediction markets, and model-based estimates | Definitions vary; participant pools are biased; good calibration on short-term questions does not guarantee accuracy on AGI |
| Scaling law continuation | Medium | Whether empirical relationships between compute/data and model performance continue to hold as we scale | Assumes continuity of current paradigms and benchmarks; may break if architectures, training regimes, or evaluation metrics change |
| Specific timeline predictions | Low | Point estimates from various methods | No track record; high parameter sensitivity |
| Economic impact forecasts | Low | Gross domestic product and labor market projections | No interdisciplinary consensus; diffusion dynamics unclear |

SOURCES: Compute trends reliability assessment based on Sevilla et al., 2022, and Sevilla and Roldán, 2024. Capital investment signal discussed in industry reporting; see, for example, Fonteneau et al., 2025, for recent figures. Timeline prediction limitations documented in Armstrong and Sotala, 2015. Timeline compression trend documented across Grace et al., 2025; Todd, 2025; and Metaculus community forecasts. Scaling law evidence from Kaplan et al., 2020, and Hoffmann et al., 2022. Economic impact forecast uncertainty discussed in Brynjolfsson, Rock, and Syverson, 2021; see also Humlum and Vestergaard, 2024, for evidence that aggregate labor market effects remain small. Robustness ratings assess the consistency and independence of each signal, not its predictive accuracy for specific timelines.

---

[2]  Some prominent forecasters have revised timelines further into the future in late 2025—notably, Daniel Kokotajlo (the lead author of AI 2027) shifting from ~2027 to ~2030 and Andrej Karpathy describing AGI as "a decade away" (Lifland, Kokotajlo, and Halstead, 2026; Patel, 2025). These revisions are healthy signs of a forecasting community that updates on evidence rather than anchoring to prior positions. However, they do not reverse the secular compression documented here: Kokotajlo's 2030 estimate remains dramatically shorter than the 2050 or later medians common five years ago, and Karpathy's "decade" timeline is more aggressive than mainstream expert surveys from the 2010s. For decisionmakers, the relevant signal is not whether AGI arrives in 2027 versus 2030—a distinction of limited operational consequence—but that the distribution of informed estimates has shifted decisively toward nearer-term scenarios over the past five years.

# Chapter 5 References

# Strategic Implications: Decisionmaking Under Uncertainty

Given the shift toward shorter timeline forecasts documented in earlier chapters and the limitations of forecasting methods, how should decisionmakers respond? This chapter outlines a framework for decisionmaking that does not require resolving the uncertainty about AGI timing. The approach positions the United States to capture the benefits of AI leadership while building the institutional capacity to manage emerging risks.

## The Asymmetry of Costs

A common argument holds that decisionmakers should plan for near-term AGI because the costs of underpreparation exceed the costs of overpreparation.

**The case for prioritizing short timelines.** If transformative AI arrives in the early 2030s without adequate preparation, the consequences could be severe: dangerous capabilities deployed without safeguards, safety failures in high-stakes domains, regulatory chaos, and loss of competitive position to nations with more coherent strategies. These harms are vivid and attributable; if something goes wrong, the failure to prepare will be evident. Conversely, if transformative AI does not arrive until 2050 but preparation began in the 2020s, the apparent downside is early investment in evaluation infrastructure, technical talent, and monitoring capacity—assets that strengthen U.S. AI leadership regardless of which timeline materializes.

**The case against over-weighting short timelines.** Excessive precaution has opportunity costs: Resources devoted to speculative scenarios are unavailable for addressing present challenges or accelerating beneficial applications. Premature or poorly designed regulation poses risks that are real but difficult to quantify: compliance costs that favor well-resourced incumbents over startups, barriers to open-source development that concentrate power among a few frontier laboratories, constraints that cripple defensive applications of current AI tools, and chilling effects on research that, by their nature, cannot be measured—one cannot count the innovations that were never pursued. These costs are often diffuse and delayed, making them easier to underweight relative to vivid near-term harms. The "asymmetry" framing can also justify almost any level of precautionary spending if taken to extremes; every marginal safety investment can be defended by gesturing at catastrophic tail risks. These tensions are mirrored in the geopolitical domain, in which analysts debate whether the greatest risks stem from the ambiguous pre-AGI period or from the competitive dynamics of the race itself (Mitre et al., 2025).

The resolution lies not in choosing between these positions but in identifying actions that provide value across scenarios, seeking strategies that perform reasonably well across multiple plausible futures. For government, this means maintaining U.S. competitive advantage; for investors, managing portfolio risk across timeline scenarios; for frontier laboratories, balancing capability development with safety investment; for research funders, allocating resources to reduce the uncertainties that matter most.

## Policies, by Timeline Sensitivity

Policies can be categorized by their sensitivity to timeline assumptions, as follows.

**Robust-across-scenarios investments** represent one such policy, whether transformative AI arrives in 2030, in 2050, or via a paradigm fundamentally different from current approaches. For government, such investments include deep technical expertise, world-class capability to evaluate AI systems, mechanisms for tracking capability trends, and forums for international coordination. These are investments in institutional capacities, the expertise to evaluate systems, the infrastructure to track trends, and forums to coordinate rather than bets on specific technical architectures, which is what makes them durable across paradigm shifts. For investors, the relevant question is less about individual portfolio hedging than whether the broader investment ecosystem maintains sufficient diversity of exposure across timeline scenarios. For frontier laboratories, these investments include safety research and evaluation capacity that provides value regardless of which capabilities emerge when. For research funders, they include investments in forecasting methodology and capability measurement that reduce uncertainty. Such investments should be priorities regardless of timeline beliefs.

**Timeline-contingent strategies** make sense only under specific timeline assumptions. Aggressive, near-term safety requirements assume short timelines; a wait-and-see approach to workforce adjustment assumes longer timelines. Decisionmakers should be explicit about these dependencies and build in mechanisms for reassessment as evidence accumulates.

A natural question arises: If the appropriate strategy emphasizes adaptive capacity over betting on specific timelines, what is the marginal value of better forecasting? The answer is that forecasting infrastructure serves the adaptive strategy itself. Better leading indicators tell decisionmakers what to watch; more-robust benchmarks distinguish genuine capability gains from measurement artifacts; tracking capability trajectories provides the evidentiary basis for reassessment triggers. The goal is not to predict with precision but to reduce the risk of surprise, inform when to revisit timeline-contingent policies, and help distinguish investments that are robust across scenarios from those that depend on specific assumptions materializing.

## Distinguishing Strategy and Policy Triggers

As discussed in Chapter 2, technical capability demonstration, operational deployment, and societal transformation may occur on different timelines. Strategy and policy should be keyed to the appropriate trigger.

**Security and safety policies should be informed by capability demonstrations.** Measures that address national security risks and dangerous capabilities should be informed primarily by demonstrated capabilities rather than widespread deployment, though operational context matters; whether a new capability actually shifts attacker-defender balances depends on real-world barriers to exploitation in addition to technical potential (Narayanan and Kapoor, 2025). A model that can assist with designing dangerous pathogens is concerning when it exists, not when it is commercially available. This logic underlies the capability-based safety frameworks adopted by frontier laboratories, which trigger enhanced protocols based on specific dangerous capabilities rather than deployment milestones. For decisionmakers, the implication is that security-relevant evaluation and monitoring must occur pre-deployment, requiring trusted access to frontier systems and the technical capacity to assess them. These policies should assume the shorter of plausible timelines, because the cost of being caught unprepared by a dangerous capability substantially exceeds the cost of building evaluation infrastructure that proves, in retrospect, to have been premature.

**Innovation and competitiveness policies should proceed independently of timeline assumptions.** Support for domestic AI research, infrastructure investment, and talent development should proceed continuously regardless of timeline beliefs. These investments compound over time: Research programs take years to mature, talent pipelines require sustained cultivation, and computing infrastructure involves long procurement and construction cycles. Starting and stopping based on shifting timeline estimates is costly and risks ceding ground to competitors with more consistent strategies. Rather than betting on a specific timeline, the aim is to ensure that U.S. institutions are positioned for leadership across the range of plausible futures and that government has sufficient technical expertise to evaluate developments, partner effectively with industry, and make informed decisions as the landscape evolves.

**Economic adjustment policies should be informed by observable economic effects.** Policies on labor markets, education, and workforce development can key their *activation* to observable economic effects rather than speculative capability forecasts; labor displacement, wage effects, and productivity shifts provide measurable signals that allow calibrated response. This graduated approach assumes that disruption unfolds over years, allowing time to observe and react. Under scenarios in which AI rapidly automates a broad swath of cognitive labor, displacement could be sudden and widespread enough that traditional adjustment mechanisms—retraining, job transitions, new sector growth—are insufficient, because comparable new employment may not emerge on any relevant timescale. However, the *infrastructure* for responding must be built in advance. Fiscal space for enhanced unemployment insurance, retraining program capacity, and portable benefit systems requires years of legislative and administrative work. Waiting for displacement to materialize before building shock absorbers would leave workers unprotected during the critical transition period. Moreover, sustained political support for innovation and competitiveness investments depends on credible commitments to managing disruption. An AI policy that prioritizes frontier capabilities while neglecting the labor market risks eroding the domestic coalition for U.S. AI leadership.

In practice, these categories are not cleanly separable. A capability demonstration may simultaneously raise security concerns and shift investment priorities; deployment decisions may be shaped by anticipated societal transformation effects. The distinction is nonetheless useful for matching the tempo and character of response to the nature of the development.

## Indicators for Ongoing Assessment

Given the limitations of existing forecasting methods, decisionmakers benefit from tracking specific indicators that illuminate the trajectory of AI capabilities. The framework defined in Table 6.1 connects observable developments to the key empirical questions—or cruxes—identified in Chapter 4. Each indicator domain maps directly onto those cruxes and thus helps decisionmakers track which of the stylized positions is gaining empirical support over time.

It is tempting to frame these indicators as "trip wires" that would automatically trigger strategic responses. Experience suggests caution about such framing. AI systems have already crossed thresholds that would have seemed remarkable just years ago—passing professional examinations, saturating benchmarks designed to remain challenging for years, achieving rapid consumer adoption. Policy did not change dramatically in response to any single capability milestone. This approach reflects, in part, the reality that capability thresholds are difficult to specify in advance, their significance is contested in real time, and appropriate responses depend on context that cannot be fully anticipated. The value of tracking these indicators lies in structuring ongoing assessment and building shared vocabulary, not in pre-committing to specific responses.

These indicators are organized around decision-relevant domains—what decisionmakers can observe and act on—rather than the cognitive taxonomies used in formal AGI frameworks. Google DeepMind's "Levels of AGI" framework classifies systems by performance depth, generality breadth, and autonomy level (Morris et al., 2024). Hendrycks and colleagues have proposed grounding AGI evaluation in psychometric theory, mapping capabilities to such domains as fluid reasoning and long-term memory (Hendrycks et al., 2025). A full crosswalk between these frameworks and decision-relevant indicators would help clarify which observ-

**TABLE 6.1**

**Capability Trajectory Indicators**

| Domain | Definition | Current Status | Indicator to Monitor | What Progress Would Signal |
|---|---|---|---|---|
| Fluid intelligence | Capacity for novel reasoning, abstraction, and problem-solving on tasks unlike training data | Frontier models match human expert baselines on some memorization-resistant benchmarks (e.g., GPQA Diamond) but remain well below expert level on more-novel tasks, such as ARC-AGI-2 and FrontierMath. | Performance on reasoning benchmarks specifically designed to test generalization over memorization | The sufficiency crux: Evidence bearing on whether current approaches can achieve genuine reasoning or face fundamental limits |
| R&D automation | Degree to which AI systems perform tasks contributing to AI R&D | AI tools are reported to substantially boost programming and machine learning research productivity (Anthropic, 2026), though humans still direct frontier research strategy and novel hypothesis generation. | Share of AI research and engineering tasks substantially automated; AI contribution to novel research directions | The takeoff crux: Evidence bearing on whether AI-accelerated AI development could compress future timelines |
| Economic utility | Measurable productivity gains and economic value from AI deployment | AI tools improve some worker productivity; aggregate economic effects are not yet visible in statistics. | Productivity effects across worker skill levels; adoption depth beyond initial pilots; aggregate productivity statistics | The diffusion crux: Evidence bearing on the gap between capability demonstration and economic transformation |
| Compute investment | Scale and growth rate of computational resources devoted to training frontier models | Frontier training runs are at the $100 million to $1 billion scale, with continued scaling investment. | Training run costs; whether capability continues scaling with compute at higher investment levels | The scaling crux: Evidence bearing on whether current approaches face diminishing returns at extreme scale |
| Sensitive capabilities | Performance on tasks with national security implications, including cybersecurity, biosecurity, and autonomous operations | Models have broad knowledge but limited ability to execute complex operational tasks autonomously. | Performance on evaluations of capabilities relevant to national security, as assessed by appropriate agencies | Security considerations: Evidence bearing on transition from theoretical knowledge to operational capability in sensitive domains |

able milestones correspond to which capability thresholds—an important direction for future work. Rather than setting thresholds that trigger automatic responses, the value in this framework is in providing a structured basis for ongoing situational awareness. It connects observable developments to the underlying questions that drive forecasting disagreement, helping decisionmakers interpret new information as it emerges.

## Summary: Principles for Strategy Under Uncertainty

The analysis in this report suggests the following principles for decisionmakers navigating uncertainty about AI trajectories:

- **Treat the shift toward shorter timelines as a real signal.** Multiple independent methods show forecasts moving earlier, reflecting observable trends in compute, investment, and capability. The range of estimates has also narrowed, with the spread between optimistic and pessimistic projections converging. This does not mean that transformative AI is imminent, but it does mean that the possibility warrants serious contingency planning.

- **Acknowledge irreducible uncertainty.** Reasonable experts disagree on substantive grounds. Waiting for consensus is not viable because disagreement may persist until developments resolve the underlying empirical questions.
- **Prioritize investments that provide value across scenarios.** For government, this means building technical expertise, evaluation capability, and monitoring capacity, as well as building the institutional capacity to assess developments as they occur and respond effectively to a variety of futures. For laboratories, it means safety and evaluation infrastructure that scales with capability. For funders, it means research that reduces decision-relevant uncertainty. Recognize that, under the shortest credible timelines—some frontier lab leaders forecast human-level AI by 2026–2029—many of the preparations described in this report may not mature in time. This argues for beginning immediately and prioritizing actions that can be executed quickly.
- **Match response timing to domain.** Security measures should respond to capability demonstrations; innovation support should proceed continuously; economic adjustment can respond to observed transformation. Different timelines apply to different strategic and policy domains.

Some robust investments—especially evaluation standards, monitoring infrastructure, and safety research—are particularly valuable when coordinated internationally (e.g., through Organisation for Economic Co-operation and Development or G7 processes); Chapter 7's agenda is not U.S.-specific.

## Chapter 6 References

# Building Better Forecasting Infrastructure

The validation challenges documented in the previous chapters are not insurmountable. Targeted research investments could substantially improve the quality of information available to decisionmakers. This chapter outlines priority areas and is structured as a request for proposals. Different actors have distinct roles to play, as follows:

- **U.S. government agencies** (National Institute of Standards and Technology [including its Center for AI Standards and Innovation], National Science Foundation, Intelligence Advanced Research Projects Agency [IARPA], Office of Science and Technology Policy, Department of Energy) can fund research infrastructure, convene stakeholders, and build internal technical capacity for ongoing assessment.
- **Private philanthropy and international bodies** can support independent scholarship that complements government priorities or fills gaps where there is underinvestment by governments.
- **Academic researchers** (computer scientists, economists, cognitive scientists, historians of technology) can bring methodological diversity and critical scrutiny to forecasting models.
- **Frontier AI laboratories** can contribute proprietary data, participate in pre-deployment evaluation, and develop internal monitoring systems designed for eventual information-sharing.
- **Forecasting platforms and research organizations** can expand tracking infrastructure and develop new metrics.

A caveat is in order. This report attempts to synthesize and summarize a spectrum of beliefs about AGI timelines, some of which are in direct contradiction with one another. Not all of the research priorities that follow can be simultaneously important; their relevance depends on which empirical assumptions prove correct. Under the shortest timelines, most of them will not matter; the capabilities they aim to measure or the institutions they aim to build would be overtaken by the developments themselves. The agenda is nonetheless presented in full because the uncertainty is genuine and because the cost of building forecasting infrastructure that turns out to be unnecessary is far lower than the cost of neglecting it.

## Adversarial Validation of Forecasting Models

**The gap:** Influential forecasting models are built by small teams and lack systematic external review. The Biological Anchors framework, for instance, represents one of the most detailed attempts to forecast transformative AI timelines but is rather complex and difficult to explain. In comparison with climate modeling, which has been developed by large expert communities over decades, compute-centric forecasting methodologies lack institutional depth (Karnofsky, 2021).

**The objective:** Establish an institutionalized adversarial review of compute-centric models and other forecasting frameworks.

**Priority areas:**

- **Independent replication of Biological Anchors and similar models, with sensitivity analysis on key parameters.** Cotra's model depends critically on assumptions about willingness to spend, compute requirements, and algorithmic progress (Cotra, 2020; Cotra, 2022). Sensitivity analysis could reveal which parameters most influence timeline estimates (Filan, 2021).
- **Development of competing models with different structural assumptions.** The forecasting ecosystem relies heavily on a small number of compute-centric frameworks; methodological diversity would help identify whether apparent convergence reflects genuine signal or shared blind spots (see Tetlock, 2005, on shared biases among expert communities).
- **"Red-team" exercises tasked with identifying failure modes in leading forecasts**, including assessing whether assumptions are reasonable and actively constructing scenarios under which predictions break down. Similar adversarial review processes and red-teaming have proven valuable in intelligence analysis (Central Intelligence Agency, 2009; National Research Council, 2011).
- **Systematic cross-model documentation of assumptions.** A comparative resource documenting parameter choices across leading forecasting frameworks (e.g., Biological Anchors or Davidson's take-off model) would enable researchers to identify where assumption differences drive divergent predictions and facilitate independent sensitivity analysis.

## Intermediate Milestone Forecasting

**The gap:** Forecasters cannot be calibrated on AGI directly. Accuracy must be tracked on intermediate, resolvable questions. The IARPA-sponsored forecasting tournament (2011–2015) demonstrated that systematic tracking of forecaster accuracy can identify individuals and methods that substantially outperform baselines (Tetlock and Gardner, 2015; Mellers et al., 2014). Greater weight could then be given to predictions from overperforming forecasters.

**The objective:** Launch high-frequency forecasting tournaments focused on six-to-24-month AI milestones, with systematic tracking of forecaster accuracy. Metaculus operates an AI Forecasting Benchmark tournament that explicitly benchmarks AI systems against human community predictions, with substantial quarterly prize pools incentivizing participation (Metaculus, undated). These tournaments can be useful in calibrating both AI and human forecasts.

**Priority areas:**

- **Identify leading indicators that predict farther-term capability trajectories.** Steinhardt's experiment tracking AI benchmark forecasts (2021–2023) found that both AI experts and superforecasters systematically underestimated progress on such benchmarks as MATH, with true performance falling in "the far right tail of their predicted distributions" (Steinhardt, 2023).
- **Track calibration of different forecaster types (machine learning researchers, superforecasters, prediction markets) on near-term questions.** In Steinhardt's experiment, the Metaculus crowd forecast and Steinhardt's machine learning–informed predictions were best calibrated, whereas Hypermind superforecasters underestimated progress by the widest margins. By the second year, machine learning researchers had improved their calibration, but superforecasters continued to underestimate benchmark progress (Steinhardt, 2023). Understanding which forecaster types best predict AI progress is a key input for weighting longer-horizon forecasts.
- **Develop methodology for aggregating forecaster performance into credibility weights for longer-term predictions.** Research on optimal forecast aggregation suggests that differential weighting based on track records can substantially improve accuracy (Mellers et al., 2014), but extending these methods to long-horizon questions with different resolution characteristics remains an open problem.

- **Create standardized question formats and resolution criteria that reduce framing effects and improve comparability.** Alternative question wordings can shift median estimates by decades (Grace et al., 2018), and differing operationalizations of AGI across platforms make cross-platform comparison difficult (Wynroe, Atkinson, and Sevilla, 2023).

## Robust Capability Measurement

**The gap:** Benchmarks degrade faster than they can be replaced. Saturation, contamination, and optimization pressure are not occasional failures but structural dynamics that will undermine any static evaluation. Organizations including METR, Epoch AI, the United Kingdom's AI Security Institute, and academic groups are working on this problem, but efforts are fragmented and underresourced relative to the pace of frontier model development.

**The objective:** Establish durable, independent institutional capacity for continuous capability evaluation—not a fixed set of benchmarks but an ongoing function that refreshes assessments faster than they can be gamed. The Frontier Model Forum has published a taxonomy of pre-deployment safety evaluations, including benchmark evaluations, red-teaming exercises, safeguard evaluations, and uplift studies, but notes that "a robust and effective evaluation ecosystem . . . will require shared understandings of other best practices" (Frontier Model Forum, 2024).

**Priority areas:**

- **Sustain and scale independent evaluation organizations.** Groups such as METR, Epoch AI, Apollo Research, RAND, and the United Kingdom and U.S. AI security institutes perform this function but lack the resources to keep pace with frontier development. Government funders (National Science Foundation, National Institute of Standards and Technology) and philanthropy should treat continuous evaluation as critical infrastructure requiring sustained support, not one-time grants.
- **Invest in dynamic evaluation methodologies.** Procedurally generated tasks, continuously refreshed test sets, and held-out evaluations maintained under strict access controls offer more durability than static benchmarks. Promising approaches include LiveCodeBench's use of recent competitive programming problems (Jain et al., 2024) and Epoch AI's FrontierMath, which uses expert-crafted, research-level math problems designed to resist saturation far longer than conventional benchmarks (Glazer et al., 2024). The goal is not a single robust benchmark but a portfolio of approaches that degrade at different rates.
- **Develop secure evaluation protocols for pre-deployment testing.** Evaluating frontier models before release requires information security measures and trusted relationships with developers. The United Kingdom's AI Security Institute has piloted such arrangements (AI Security Institute, 2024); expanding this capacity and establishing similar functions in other jurisdictions would improve coverage.
- **Support research that decomposes capabilities into underlying cognitive domains.** Better theoretical grounding for what benchmarks measure—and how different capabilities relate to decision-relevant thresholds—would help decisionmakers interpret results. Work by Hendrycks and colleagues (2025) on mapping AI evaluation to psychometric constructs offers one potential direction.

## Economic Diffusion Modeling

**The gap:** We lack models that map from technical capability to societal transformation with validated parameters. Early efforts to model these dynamics—for example, comparing growth trajectories under assistive versus autonomous AI development scenarios—illustrate both the stakes and the difficulty of this work (Sytsma, 2025). Computer scientists are ill-equipped to predict economic diffusion. Even the first step of extrapolating from AI coding ability to developer productivity is difficult. As one study notes, "Real-world software development involves working with large, evolving codebases, collaborating across teams with different coding standards . . . . The gap between benchmark performance and comprehensive production utility across the full development lifecycle remains largely unexplored" (Kumar et al., 2025, p. 2).

    **The objective:** Build quantitative models of AI diffusion that incorporate regulatory, organizational, and infrastructure constraints. The AI Workforce Research Hub established under the AI Action Plan, which is tasked with recurring analyses and scenario planning for a range of AI impact levels, provides a natural institutional vehicle for this work.

    **Priority areas:**

-  **Track adoption metrics for existing AI tools (coding assistants, language models in enterprise) to establish empirical baselines.** Useful indicators would include enterprise deployment rates, the share of developers actively using AI assistants, and frequency of use across task types. Frontier AI labs are well-positioned to contribute these data through system cards and usage disclosures (e.g., Anthropic, 2026), which is particularly important given that productivity studies to date have largely evaluated tools well behind the capability frontier (e.g., Weisz et al., 2025).
- **Measure productivity impacts of AI deployment with rigorous methodology.** Early evidence is mixed: an IBM study ($N = 669$) found unevenly distributed productivity gains (Weisz et al., 2025), a systematic review of 37 studies documented benefits alongside concerns about cognitive off-loading and code quality (Mohamed, Assi, and Guizani, 2025), and METR's randomized controlled trial found that experienced developers were 19 percent slower with AI tools, possibly reflecting learning costs (Becker, 2025). Moreover, existing benchmarks may be too narrow to measure the full suite of capabilities that real-world development demands (Kumar et al., 2025), and existing studies largely evaluate tools behind the capability frontier, making continued measurement essential.
- **Develop sector-specific models of automation potential and adoption barriers.** Identifying disparate impacts on various sectors will allow better strategy and policy development and implementation.
- **Engage economists and organizational researchers in AI forecasting** (currently dominated by computer scientists). Approaches such as exploratory scenario modeling and assumption-based planning could supplement the analytic frameworks in use (Kilian, Ventura, and Bailey, 2023).

## Monitoring Infrastructure

**The gap:** We lack real-time indicators that would provide early warning of rapid capability gains.

    **The objective:** Develop standardized metrics that can be tracked regularly and would signal significant capability transitions.

    **Priority areas:**

- **Identify measurable proxies for AI contribution to AI research.** A July 2025 expert workshop held by the Center for Security and Emerging Technology found broad agreement that leading AI companies are already using their own systems to accelerate R&D, with each generation contributing to building the next, though participants disagreed sharply on how far this automation could ultimately go and whether it risks a rapid, difficult-to-control acceleration of capabilities (Toner et al., 2026). Measuring this contribution empirically would require tracking both task-level metrics, such as the share of code commits or experiment implementations completed by AI agents, and broader indicators of the overall pace of AI progress, such as effective compute and compute multipliers, that capture how algorithmic efficiency gains amplify raw hardware investments.

- **Develop internal, shareable models of AI R&D automation.** Frontier labs are in the best position to measure early signs of automated AI research—for example, the number and duration of experiments run by AI agents, the share of code commits or experiment implementations authored by AI, error rates, and the kinds of research problems delegated to agents. Labs should build internal monitoring models that use these metrics to estimate the degree of automation of AI R&D. In the near term, these models and underlying data could remain proprietary, but they should be designed in standardized formats so that, in future national or international coordination regimes, labs can securely share either the indicators or aggregated outputs with regulators and trusted evaluators.

- **Design standards for privacy-preserving reporting of frontier model capability evaluations.** Such standards can build on existing safety frameworks, which already commit many frontier AI companies to evaluate models for severe risks, including automated AI R&D, cybersecurity, and biological weapons (METR, undated), and to publish high-level safety frameworks under the Seoul Frontier AI Safety Commitments. Research is needed on *how* evaluation results should be summarized and shared. Key questions include the following: What minimal set of capability indicators should be reported? At what level of aggregation? How can labs provide regular updates to regulators or trusted evaluators without revealing proprietary information or operationally sensitive details?

- **Build integrated dashboards of leading indicators that go beyond raw compute and benchmark scores.** Existing efforts, such as Epoch AI's trends database, already track compute scaling, hardware efficiency, power consumption, and training costs with regular updates (Epoch AI, undated). A monitoring infrastructure for takeoff risk would extend this by (1) adding standardized indicators of AI contribution to AI R&D (for example, the share of RE-Bench–style research engineering tasks completed by AI agents), (2) summarizing labs' reported evaluations against frontier risk thresholds (e.g., automation of AI R&D, cybersecurity, development of biological weapons), and (3) where feasible, including adoption metrics, such as the extent to which autonomous agents are deployed in critical or safety-relevant workflows.

On short AGI timelines—months to a few years—AI forecasting may end up looking less like traditional technology roadmapping and more like quantitative finance. Rather than relying on a single, "best" model, we should expect a portfolio of statistical models that ingest heterogeneous, high-frequency signals (compute spending, benchmark results, internal deployment metrics, hiring and capital expenditure plans, even sentiment in technical communities) and human analysts who make adjustments based on contextual knowledge. Because many of the most informative indicators of automated AI R&D—such as experiment throughput, failure patterns, and the share of code or experiment design authored by AI agents—are proprietary, frontier labs are best positioned to run such models. A useful research direction is to design these multisignal monitoring systems now, including standardized inputs and summary statistics, so that their outputs could be securely shared with regulators or international partners if future coordination frameworks call for it.

# Chapter 7 References

# Conclusion

The forecasting landscape for AGI presents decisionmakers with a genuinely difficult challenge. Multiple signals (expert surveys, prediction markets, compute models, and capital allocation) point toward earlier estimated arrival dates, with central estimates from several methods clustering in the 2030s. Yet the methodological foundations underlying these forecasts are immature, lacking resolved predictions for calibration, robust benchmarks for measurement, and adversarial validation of models.

In this report, we do not argue that AGI will arrive by any specific date. The evidence does not support such confidence. Rather, we argue the following:

1. **The shift to earlier predictions of timelines to AGI is real and consistent.** Multiple independent methods show movement toward earlier expectations, reflecting observable scaling of compute and capital, combined with empirical progress on benchmarks.

2. **Reasonable experts disagree on substantive grounds.** Proponents and skeptics of near-term AGI both hold defensible positions about the strengths and limitations of existing approaches. The uncertainty is genuine, not a failure of analysis.

3. **Waiting for consensus is not a strategy.** Expert disagreement is likely to persist until AGI either arrives or demonstrably fails to materialize. Decisionmakers must act under uncertainty.

4. **Robust investments dominate speculative bets.** Strategies that build capacity in the United States—technical talent, evaluation infrastructure, monitoring systems, international coordination—provide value across scenarios and should be priorities regardless of timeline beliefs.

The forecasting community has a critical role to play. By investing in the research agenda outlined in Chapter 7—adversarial validation, intermediate milestone tracking, robust measurement, and monitoring infrastructure—the forecasting community can provide decisionmakers with better instruments for navigating this uncertainty.

As AI capabilities advance, the quality of our predictions becomes increasingly consequential. The goal is to directly engage with uncertainty, characterize it honestly, identify where confidence is warranted, and build the institutional capacity to respond to a variety of futures.

# Evolution of AI Forecasting Approaches

The practice of forecasting AGI has evolved from philosophical speculation into a discipline attempting to ground itself in empirical data. This appendix surveys the historical development of these methods, providing context for why current forecasts are shifting rapidly.

## Early Forecasting and Hardware Extrapolation (1950s–2010s)

Early AI forecasting revolved around conceptual and hardware-centric arguments. Alan Turing's "Imitation Game" framed machine intelligence in behavioral terms, implicitly suggesting a benchmark for human-level AI (Turing, 1950). The Dartmouth Workshop marked a period of optimism about AI's near-term potential (McCarthy et al., 2006), though the degree of early hype is commonly exaggerated; it was concentrated among a few prominent voices, while others, including Claude Shannon, expressed considerably more caution (Muehlhauser, 2016). Still, the field's most visible predictions systematically underestimated the complexity of perception, commonsense reasoning, and the software engineering required to replicate these traits.

As Moore's Law became observable, forecasters such as Hans Moravec and Ray Kurzweil introduced "hardware parity" arguments, positing that, once commodity hardware matched the theoretical computational capacity of the human brain, human-level AI would follow (Moravec, 1998). Kurzweil's Law of Accelerating Returns projected accelerated improvements to predict a "Singularity" in the mid-21st century (Kurzweil, 2005). Retrospective analysis suggests that algorithms, data, and training methodology appear to matter independently, beyond matching hardware capacity (Armstrong, Sotala, and Ó hÉigeartaigh, 2014; Cotra, 2020; Karnofsky, 2021).

The 2000s and 2010s saw institutionalization of forecasting. Anders Sandberg and Nick Bostrom outlined whole brain emulation (WBE) as a pathway to AGI (Sandberg and Bostrom, 2008), while Bostrom later systematized multiple pathways, including "de novo" AI (Bostrom, 2014). Early expert surveys began to quantify the field's beliefs, revealing a persistent 15-to-25-year horizon bias—a tendency for experts to predict AGI as being roughly two decades away, regardless of when asked (Armstrong, Sotala, and Ó hÉigeartaigh, 2014). A 2023 expert survey showed a notable 13-year shift in median timelines compared with 2022 (Grace et al., 2025), though the resulting median of 2047 still falls within the characteristic two-decade window, leaving open the question of whether empirical progress has genuinely broken the pattern or merely shifted it.

## Crowd Forecasting and Prediction Markets

Crowd forecasting relies on aggregating diverse judgments to cancel individual biases. Platforms such as Metaculus and Manifold use scoring rules or market mechanisms to incentivize accuracy. Metaculus has documented significant shifts in AGI median estimates, moving from 50 years away in 2020 to under ten years by 2025, though these estimates use platform-specific resolution criteria that set a lower bar than many common definitions of AGI (Metaculus, undated; Todd, 2025). These platforms show strong calibration on short-term, well-specified questions, but it is unclear whether that calibration transfers to long-horizon, fuzzy concepts, such as AGI, for which resolution criteria are debatable.

Distinct from open prediction markets, elite forecaster aggregation offers a complementary approach. One example is Samotsvety Forecasting, a small team of forecasters selected for strong performance on Metaculus and other forecasting platforms who aggregate members' views by trimming extremes and taking the geometric mean of odds (Samotsvety Forecasting, 2023). In a January 2023 update prepared for Epoch AI's literature review, the team's aggregate implied about a 28 percent probability of AGI or transformative AI by 2030 and a median arrival date around 2043—toward the short end of expert and community judgment-based forecasts but broadly in the same range as other aggressive aggregates (Samotsvety Forecasting, 2023; Wynroe, Atkinson, and Sevilla, 2023). Epoch AI's review rated the Samotsvety timelines as the most methodologically credible of the judgment-based forecasts it considered but emphasized that substantial uncertainty and model dependence remain (Wynroe, Atkinson, and Sevilla, 2023).

An emerging methodology uses LLMs themselves as forecasters. ForecastBench, developed by the Forecasting Research Institute, benchmarks LLM forecasting ability against human superforecasters across a broad variety of questions spanning politics, finance, and other domains (Karger et al., 2024). Bastani, Kučinskas, and Karger (2025) use a simple, linear extrapolation of improving state-of-the-art LLM performance on the benchmark to project LLM-superforecaster parity around late 2026, with a reported 95 percent confidence interval from December 2025 to January 2028. If that trajectory holds, LLM-generated forecasts could soon serve as a scalable complement to elite human judgment on well-specified questions, though whether that calibration would extend to fuzzier, long-horizon questions (such as AGI timelines) remains an open question.

## Compute-Centric Models and Biological Anchors

The Biological Anchors framework, developed by Ajeya Cotra for Open Philanthropy, represents the most mechanistically rigorous approach to AGI forecasting (Cotra, 2020). It estimates training compute required for AGI by using biological systems as anchors—the computational power of the human brain, the amount of computation performed over the course of natural selection, or the amount of computation performed over a human lifetime—and combines these metrics with projections of future compute availability. The original 2020 report estimated median transformative AI around 2050; Cotra's 2022 update, incorporating faster-than-expected algorithmic efficiency improvements and scaling progress, shifted her median to around 2040 (Cotra, 2022).

These models are heavily informed by empirical scaling laws, which show that model performance scales as a power law with compute and data (Kaplan et al., 2020; Hoffmann et al., 2022). However, the models are fragile: They rely on order-of-magnitude estimates for biological computation, and modest parameter changes can shift timelines by decades (Cotra, 2020; Karnofsky, 2021). Critics note that biological anchors may not capture the relevant computation for intelligence and that algorithmic breakthroughs could render compute estimates obsolete (Alexander, 2022; Yudkowsky, 2021).

Epoch AI's "Direct Approach," methodologically distinct from Biological Anchors, uses neural network scaling laws to directly bound compute requirements for transformative AI without biological reference points (Barnett and Besiroglu, 2023). The model's central scenario combines projections of AI investment growth, hardware price-performance improvements, and algorithmic efficiency gains against an upper bound on the training compute required for transformative AI derived from neural scaling laws. This model produces a median estimate around 2033 (as of March 2026) and provides an interactive interface allowing users to adjust parameters, including investment growth, hardware efficiency, algorithmic progress, and compute requirements (Atkinson et al., 2024). The divergence between Biological Anchors and Direct

Approach estimates illustrates how different methodological choices within the compute-centric paradigm can yield substantially different timelines.

## Expert Surveys and Elicitation

Structured surveys of AI researchers reveal the distribution of beliefs within the field and track how those beliefs change. The trajectory illustrates a consistent shift toward earlier estimates: Müller and Bostrom (2016) found roughly a 50 percent probability of HLMI by midcentury; Grace et al. (2018) reported a median of 2061; and Grace et al.'s 2023 survey documented a sharp shift, with the median moving to 2047 (Grace et al., 2025). Whether this compression represents genuine updating based on empirical progress or a continuation of the historical tendency to place AGI roughly 15–25 years in the future remains an open question (Armstrong, Sotala, and Ó hÉigeartaigh, 2014).

Expert surveys are subject to methodological limitations, including framing effects, selection bias, and anchoring. The Forecasting Research Institute's Longitudinal Expert AI Panel attempts to address some limitations through repeated elicitation from a consistent panel, enabling the tracking of individual belief updates over time (Murphy et al., 2025). This longitudinal approach may help distinguish genuine updating from noise and identify which experts update most appropriately in response to new evidence.

The Delphi method—structured expert consensus through iterative, anonymous feedback—has been applied to AI forecasting across multiple domains. A 2025 review surveyed 13 Delphi studies on AI futures conducted between 2014 and 2024, covering applications in health care, manufacturing, and general capability development (Alon et al., 2025). Innovations include Real-Time AI Delphi, which integrates generative AI models to support the consensus-building process, potentially addressing traditional limitations of extended time frames and expert dropout (Calleo and Pilla, 2025).

## Scenario Analysis and Takeoff Dynamics

Scenario-based forecasting constructs multiple plausible futures rather than precise predictions. Scenarios typically distinguish between *slow takeoff* (continuous, observable progress that may still be super-exponential) and *fast takeoff* (discontinuous capability gains that outpace external observation) (Bostrom, 2014; Christiano, 2018). Tom Davidson's takeoff dynamics model provides quantitative structure to these scenarios, modeling feedback loops between AI capabilities and AI research productivity (Davidson, 2023).

Recent influential scenarios include Aschenbrenner's *Situational Awareness*, which argues for AGI by 2027–2030 via scaling plus algorithmic improvements (Aschenbrenner, 2024). AI 2027, a scenario exercise tracing a specific trajectory from current capabilities through recursive self-improvement, offers the most detailed public articulation of this position, though its authors have since noted that their median estimates have shifted later toward 2030 as new evidence has accumulated (Kokotajlo et al., 2025).

By specifying intermediate milestones, such forecasting aims to make an otherwise abstract debate more concrete: Proponents argue that it forces critics to identify which specific steps they find implausible, while skeptics contend that stringing together individually plausible steps can produce an implausible whole.

## Empirical Trend Extrapolation

Organizations such as Epoch AI and METR compile datasets on model capabilities, compute usage, and deployment to extrapolate trends (Epoch AI, 2026; Kwa et al., 2025). Epoch AI's tracking shows training compute for frontier models growing fourfold to fivefold each year, while METR's task-completion analysis finds the time horizon for 50 percent AI task completion doubling approximately every seven months (Epoch AI, 2026; Kwa et al., 2025).

However, empirical extrapolation depends on benchmark validity, data quality, and institutional independence. Benchmarks face saturation (models approaching ceiling performance), contamination (test data leaking into training sets) (Xu et al., 2024), and gaming (optimization for benchmark performance rather than underlying capabilities) (Raji et al., 2021). When empirical analysis is conducted by organizations linked to major labs or funders, potential conflicts of interest arise. The field needs clearer disclosure requirements, open data where feasible, and independent replication before policy use.

Epoch AI's GATE model represents an integration of empirical trend extrapolation with economic diffusion modeling, combining AI capability forecasting with estimates of deployment timelines and productivity impacts (Erdil et al., 2025). This synthesis approach addresses a gap identified in Chapter 5: the disconnect between technical capability forecasts and their economic implications. Such integrated models may prove increasingly valuable as decisionmakers seek to translate capability projections into actionable economic and workforce planning.

## Outside-View and Alternative Pathway Models

The forecasting methodologies covered in the previous sections predominantly employ inside-view reasoning, building models from specific technical parameters or extrapolating observable trends. This section examines complementary approaches that adopt explicit outside-view epistemologies or consider alternative routes to AGI.

### Semi-Informative Priors

The semi-informative priors framework, developed by Davidson, represents a fundamentally different approach to AGI timeline forecasting (Davidson, 2021). Rather than anchoring estimates to biological computation or scaling laws, it applies Laplace's law of succession and its variations, using the observed history of attempts to develop transformative technologies to derive base-rate probabilities for AGI, with parameters governing first-trial probability, prior beliefs about difficulty, and what counts as an independent attempt.

Depending on parameter assumptions—particularly the choice of trial definition and how rapidly AI R&D inputs are growing—the framework's central estimates range from the mid-21st century under base assumptions to as early as the 2030s when accounting for the rapid growth of computation in AI R&D. The framework generally produces longer timelines than inside-view models, though the gap narrows substantially under some configurations. Its value lies less in its specific outputs than in providing a systematic outside-view check: Whereas inside-view models may share correlated errors, semi-informative priors draw on base rates from technological development more broadly (Davidson, 2021).

### Whole Brain Emulation Roadmapping

WBE represents an alternative pathway to AGI independent of machine learning advances. The technical roadmap by Sandberg and Bostrom (2008) systematically analyzes prerequisites across three domains: scanning technology to image brain structure at sufficient resolution, translation of scanned data into functional models, and computational infrastructure for simulation. Sandberg and Bostrom (2008) estimate approximately 50 percent probability of WBE by 2064, though timelines depend critically on which neural features prove computationally essential.

Although contemporary forecasting focuses on deep learning pathways, WBE remains relevant for several reasons. It provides an alternative route with different technical bottlenecks; progress or stagnation in deep learning may not affect WBE timelines symmetrically. WBE also offers clearer measurable milestones (scanning resolution, simulation speed) compared with ambiguous capability benchmarks. A 2023 Foresight Institute workshop revisited WBE feasibility in light of recent AI advances, suggesting renewed interest in this pathway (Duettmann and Sandberg, 2023).

## Observation Selection Effects and Evolutionary Arguments

A common argument for AGI feasibility holds that, because evolution produced human intelligence, purposeful engineering should achieve similar results more efficiently. Shulman and Bostrom (2012) analyze how observation selection effects complicate this inference. The core problem is that observers necessarily find themselves on planets where intelligence evolved, regardless of how improbable that evolution was. Whether intelligence evolves on 1 in 10 or 1 in $10^{1,000}$ Earth-like planets, observers see identical evidence.

The authors explore whether this evolutionary argument can be salvaged through evolutionary timing analysis, convergent evolution (intelligence-related traits in octopuses, corvids, and elephants provide unbiased evidence), and different anthropic reasoning frameworks. They conclude that observation selection effects "do not cripple the evolutionary argument" (Shulman and Bostrom, 2012, p. 19). In terms of forecasting, this work demonstrates that even widely cited feasibility arguments require careful epistemological analysis and identifies which evolutionary observations provide genuinely unbiased information.

# Appendix References

# Abbreviations

| | |
|---|---|
| AGI | artificial general intelligence |
| AI | artificial intelligence |
| ARC-AGI | Abstraction and Reasoning Corpus |
| DARPA | Defense Advanced Research Projects Agency |
| FLOPS | floating-point operations per second |
| HLMI | high-level machine intelligence |
| IARPA | Intelligence Advanced Research Projects Activity |
| LLM | large language model |
| METR | Model Evaluation and Threat Research |
| MMLU | Massive Multitask Language Understanding |
| R&D | research and development |
| WBE | whole brain emulation |

# About the Authors

**Gopal P. Sarma** is a strategic technology leader and a member of the AI security cluster at RAND. He conducts research on policy development and portfolio management related to emerging technologies and national security. Sarma holds an M.D., Ph.D. in theoretical applied physics.

**Sunny D. Bhatt** is a senior policy analyst at RAND. He conducts research on national security, defense, and intelligence issues, including chemical, biological, radiological, nuclear, and high-yield explosives; organizational performance; and AI. Bhatt holds a master of public policy.

**Michael Jacob** is a technical resident and expert at RAND. He conducts technical and policy research on such topics as intelligence analysis and AI security. Jacob holds an M.A. in international affairs and international economics.

**Rachel Steratore** is a policy researcher at RAND. She conducts technical and policy research on how people interpret, evaluate, and trust AI and other emerging technologies, with the goal of supporting transparent, safe, and equitable innovation. Steratore holds a Ph.D. in engineering and public policy.